\documentclass[11pt,a4paper,american]{extarticle}
\usepackage{a4wide}
\usepackage[active]{srcltx}
\usepackage[export]{adjustbox}
\usepackage{amssymb}
\usepackage{amsmath}
\usepackage{amsfonts}
\usepackage{amsthm}
\usepackage{array,multirow,bigdelim}
\usepackage{authblk}
\usepackage{babel}
\usepackage{bbm}
\usepackage{bm}
\usepackage{braket}
\usepackage{cite}
\usepackage{color}
\usepackage{comment}
\usepackage{enumerate}
\usepackage{esint}
\usepackage{etoolbox}
\usepackage{float}
\usepackage[T1]{fontenc}
\usepackage{graphicx}
\usepackage{hyperref}
\usepackage[latin1]{inputenc}
\usepackage{latexsym}
\usepackage{mathrsfs}
\usepackage{mathtools}
\usepackage{relsize}
\usepackage{setspace}
\usepackage{slashed}
\usepackage{soul}
\usepackage{subcaption}
\usepackage{todonotes}
\usepackage{xcolor}

\newcommand{\xdownarrow}[1]{%
	{\left\downarrow\vbox to #1{}\right.\kern-\nulldelimiterspace}
}

\apptocmd{\sloppy}{\hbadness 10000\relax}{}{}

\def\a{\alpha}
\def\dif{\text{d}}
\newcommand{\s}{\sigma}

\graphicspath{ {Images/} }

\newcommand{\Tr}{ \mathrm{Tr}}

\newcommand{\be}{\begin{equation}\label}
\newcommand{\ee}{\end{equation}}
\newcommand{\bea}{\begin{eqnarray}\label}
\newcommand{\eea}{\end{eqnarray}}

\makeatletter
\newcommand*{\textoverline}[1]{$\overline{\hbox{#1}}\m@th$}
\newcommand*\bigcdot{\mathpalette\bigcdot@{.65}}
\newcommand*\bigcdot@[2]{\mathbin{\vcenter{\hbox{\scalebox{#2}{$\m@th#1\bullet$}}}}}
\makeatother

\newcommand{\eq}[1]{\begin{equation}#1\end{equation}}
\newcommand{\eqs}[1]{\begin{equation}\begin{split}#1\end{split}\end{equation}}
\date{}

\makeatletter

\allowdisplaybreaks
\numberwithin{equation}{section}
\author[1]{C. Armstrong\thanks{connor.armstrong@durham.ac.uk}}
\author[1]{A. Lipstein\thanks{arthur.lipstein@durham.ac.uk}}
\author[1]{J. Mei\thanks{jiajie.mei@durham.ac.uk}}

\affil[1]{Department of Mathematical Sciences, Durham University,
\authorcr  Stockton Road, DH1 3LE, Durham, United Kingdom}

\makeatother

\begin{document}

\title{Enhanced Soft Limits in de Sitter Space}

\maketitle

\begin{abstract}
In flat space, the scattering amplitudes of certain scalar effective field theories exhibit enhanced soft limits due to the presence of hidden symmetries. In this paper, we show that this phenomenon extends to wavefunction coefficients in de Sitter space. Using a representation in terms of boundary conformal generators acting on contact diagrams, we find that imposing enhanced soft limits fixes the masses and four-point couplings (including curvature corrections) in agreement with Lagrangians recently derived from hidden symmetries. Higher-point couplings can then be fixed using a bootstrap procedure which we illustrate at six points. We also discuss implications for the double copy in de Sitter space.
\end{abstract}

\pagebreak

\tableofcontents

\section{Introduction}
There is a deep relation between soft limits of scatteing aplitudes and hidden symmetries. For example, the soft theorems of graviton amplitudes \cite{Weinberg:1965nx,White:2011yy,Cachazo:2014fwa} encode extended BMS symmetry \cite{Strominger:2013jfa,He:2014laa}, while soft limits of pion amplitudes encode spontaneously broken chiral symmetry of QCD \cite{Arkani-Hamed:2008owk}. Pions are the Goldstone bosons associated with spontaneous symmetry breaking and are described by a low-energy effective action known as the non-linear sigma model (NLSM) \cite{Gell-Mann:1960mvl,Weinberg:1968de,Weinberg:1978kz}. Of particular interest for this paper is a property of NLSM amplitudes known as the Adler zero \cite{Adler:1964um}, which is an example of an enhanced soft limit. A scattering amplitude is said to exhibit an enhanced soft limit when it scales like $\mathcal{O}(p^\sigma)$, where $p$ is the soft momentum and $\sigma$ is an integer greater than the expectation based on counting the number of derivatives per field in the Lagrangian. For the NLSM, $\sigma=1$. More generally, $\sigma$ can be no higher than three and the cases $\sigma=2,3$ correspond to the Dirac-Born-Infeld (DBI) and special Galileon theories, respectively \cite{Cheung:2014dqa,Cheung:2016drk}. Enhanced soft limits arise from cancellations among Feynman diagrams of different topology and are a consequence of symmetries \cite{Hinterbichler:2015pqa,Cheung:2016drk}. In the NLSM, this is just an ordinary shift symmetry but in the other two cases the symmetries are higher shift symmetries which are nontrivially realised from the point of view of the Lagrangian and are often referred to as hidden symmetries.

Soft limits also play an important role in cosmology. For example in the context of inflation, where the early universe is approximately described by de Sitter space (dS), they provide constraints relating higher-point correlators to conformal transformations of lower-point correlators \cite{Maldacena:2002vr,Creminelli:2012ed,Hinterbichler:2013dpa}, and certain inflationary 3-point functions can be deduced from soft limits of 4-point de Sitter correlators \cite{Creminelli:2003iq,Assassi:2012zq,Kundu:2014gxa,Kundu:2015xta,Shukla:2016bnu}. Lagrangians for DBI and sGal theories were also recently deduced from higher shift symmetries in dS \cite{Bonifacio:2021mrf}. These Lagrangians have nontrivial masses and curvature corrections away from the flat space limit. As we will see in this paper, the NLSM can be trivially uplifted to dS space since curvature corrections would break the shift symmetry. It is therefore natural to ask if the wavefunction coefficients of these theories (which can be computed from Witten diagrams ending on the future boundary of dS \cite{Maldacena:2002vr,McFadden:2009fg,McFadden:2010vh,Maldacena:2011nz,Ghosh:2014kba}) exhibit enhanced soft limits analogous to their scattering amplitudes in the flat space limit. The goal of this work is to provide evidence that this is indeed the case. Wavefunction coefficients for the NLSM, DBI, and sGal theories were previously studied in  flat space, but they do not exhibit enhanced soft limits in this background \cite{Bittermann:2022nfh}.

To study soft limits of wavefunction coefficients, it is natural to work in dS momentum space \cite{Raju:2011mp,Raju:2012zr,Bzowski:2013sza}, which is also the standard language used for cosmology (see \cite{Maldacena:2011nz,Arkani-Hamed:2015bza,Arkani-Hamed:2017fdk,Arkani-Hamed:2018kmz,Sleight:2019hfp,Baumann:2020dch,Pajer:2020wxk,Goodhew:2020hob,Meltzer:2020qbr,Bzowski:2020kfw,Goodhew:2021oqg,Meltzer:2021zin,Sleight:2021plv,Pimentel:2022fsc,Jazayeri:2022kjy,Bzowski:2022rlz,Armstrong:2022jsa} for some recent developments). Another technique we will employ is to express the wavefunction coefficients in terms of boundary conformal generators acting on contact diagrams \cite{Eberhardt:2020ewh,Roehrig:2020kck,Gomez:2021qfd,Gomez:2021ujt,Diwakar:2021juk,Herderschee:2022ntr,Herderschee:2021jbi,Armstrong:2022csc}. Soft limits can then computed by Taylor expanding bulk-to-boundary propagators in the contact diagram, acting on them with boundary conformal generators, and using the equations of motion to remove terms which are not linearly independent. Starting with a general effective action with unfixed masses and couplings (including curvature corrections), we then find that imposing enhanced soft limits of the tree-level 4-point wavefunction coefficients fixes all the masses and 4-point couplings for the DBI and sGal theories in agreement with the Lagrangians constructed in \cite{Bonifacio:2021mrf}. For the NLSM, we find that enhanced soft limits forbid mass terms or curvature corrections, so the Lagrangian can be trivially lifted from flat space. These results in turn allow us to fix all the parameters of the generalised double copy prescription proposed in \cite{Armstrong:2022csc}, which relates the 4-point tree-level wavefunction coefficient of the NLSM model to those of the DBI and sGal theories \footnote{The double copy was first proposed in the context of scattering amplitudes, relating graviton amplitudes to the square of gluon amplitudes \cite{Bern:2008qj,Bern:2010ue}. For recent work on the double copy for (A)dS correlators see for example \cite{Farrow:2018yni,Lipstein:2019mpu,Alday:2021odx,Jain:2021qcl,Zhou:2021gnu,Sivaramakrishnan:2021srm,Cheung:2022pdk,Armstrong:2020woi,Albayrak:2020fyp,Herderschee:2022ntr,Drummond:2022dxd,Alday:2022lkk,Bissi:2022wuh}.}.  Above four points, there must be non-trivial cancellations between contact and exchange Witten diagrams in order to have enhanced soft limits. Since lower-point couplings feed into the exchange diagrams, in principle this allows us to fix all higher-point couplings using a bootstrap procedure, which we demonstrate for the NLSM and DBI theory at six points. The method can also be applied to the sGal theory above four points, but the Witten diagrams become very numerous so we save that for future work.

This paper is organised as follows. In section \ref{review}, we review enhanced soft limits and their underlying symmetries in the context of scattering amplitudes, the Lagrangians for the NLSM, DBI, and sGal theories in dS, and the method for computing wavefunction coefficients in terms of boundary differential operators acting on contact diagrams. In section \ref{fourpoint}, we use enhanced soft limits to fix the masses and 4-point couplings of the NLSM, DBI, and sGal theories in dS, and comment on the the double copy of 4-point wavefunction coefficients. In section \ref{6pt}, we describe the extension of this procedure to higher points and use it to fix all 6-point couplings of the NLSM and DBI theory. We then present our conclusions and future directions in section \ref{sec:conclusion}. There are also three Appendices containing details about our four and six-point calculations.

\section{Review} \label{review}

In this section we will briefly review enhanced soft limits and their relation to shift symmetries in flat space, mainly following \cite{Cheung:2016drk}. We then review the Lagrangians for the NLSM, DBI, and sGal theories in dS and explain how to compute cosmological wavefunction coefficients.

\subsection{Symmetries and Soft Limits}

Let us first consider a
scalar field theory with the following global shift symmetry:
\begin{equation}
\delta \phi = \theta.\label{shift}
\end{equation}
The field $\phi$ is a Goldstone boson and can be produced from the
vacuum by acting with the Noether current associated
with the shift symmetry: 
\begin{equation}
\left\langle \phi(p)\right|J^{\mu}(x)\left|0\right\rangle =ip^{\mu}Fe^{ip\cdot x},
\end{equation}
where the right-hand-side is fixed by Lorentz invariance up to an overall constant
$F$. Inserting the current between incoming and outgoing states
then gives 
\begin{equation}
\left\langle out\right|J^{\mu}(0)\left|in\right\rangle =-\frac{p^{\mu}}{p^{2}}F\left\langle out+\phi(p)\right|\left.in\right\rangle +R^{\mu}(p),
\end{equation}
where $p^\mu$ is the difference between the momenta of the in and out states. The first term on the right hand side contains a pole for the emission of a Goldstone
boson and we assume that $p\cdot R$ vanishes as $p^{\mu}\rightarrow0$ (which requires the absence of cubic vertices involving the Goldstone boson).
Multiplying by $p^{\mu}$ and noting that the current is conserved
up to contact terms which do not contribute to the scattering amplitude
after LSZ reduction, we find that 
\begin{equation}
\left\langle out+\phi(p)\right|\left.in\right\rangle =\frac{1}{F}p\cdot R.
\end{equation}
From this we immediately see that the amplitude for $\phi$ production
vanishes in the soft limit: 
\begin{equation}
\lim_{p\rightarrow0}\left\langle out+\phi(p)\right|\left.in\right\rangle =\mathcal{O}(p).
\end{equation}
This is the famous Adler zero \cite{Adler:1964um}.

Now let's consider a scalar theory with a higher shift symmetry:
\begin{equation}
\delta\phi=\theta_{\mu_{1}...\mu_{k}}x^{\mu_{1}}...x^{\mu_{k}}+...,
\label{highershift}
\end{equation}
where $\theta$ is a constant and the ellipsis denote field-dependent
terms that we will not need to consider. This
can be thought of as a special case of the shift in \eqref{shift} after
promoting $\theta$ to a function of position, which is the standard
way to compute the Noether current. As a result, one finds that the
Noether current associated with \eqref{highershift} is roughly speaking obtained by dressing the Noether current associated with \eqref{shift} with $x^{\mu_{1}}...x^{\mu_{k}}$. Inserting the new current between in and out states and repeating
similar steps to the argument above, one then finds that the soft
limit of $\left\langle out+\phi(p)\right|\left.in\right\rangle $
vanishes after being acted on by $k$ momentum derivatives which arise
from Fourier transforming $x^{\mu_{1}}...x^{\mu_{k}}$ to momentum
space, implying a higher-order Adler zero:
\begin{equation}
\lim_{p\rightarrow0}\left\langle out+\phi(p)\right|\left.in\right\rangle =\mathcal{O}\left(p^{k+1}\right).\label{enhanced}
\end{equation}
The NLSM, DBI, and sGal theories correspond to $k=0,1,2$, respectively. This behaviour arises from nontrivial cancellations among
Feynman diagrams and is therefore referred to as an enhanced soft limit. 

\subsection{de Sitter Lagrangians}

It is easy to write down the Lagrangian for the NLSM in dS$_{d+1}$:
\begin{equation}
\frac{\mathcal{L}_{\mathrm{NLSM}}}{\sqrt{-g}}={\rm Tr}\left(\partial_{\mu}U^{\dagger}\partial^{\mu}U\right),\,\,\,U=\exp\left(i\phi\right),
\end{equation}
where $\phi$ is in the adjoint of an $SU(N)$ flavour symmetry. No masses or curvature
corrections are allowed because they would spoil the shift symmetry
in \eqref{shift}. Later on we will deduce this fact from enhanced soft limits of the wavefunction coefficients. In practice, it is also convenient to use the parametrisation $U =(\mathbb{I}+\Phi)(\mathbb{I}-\Phi)^{-1}$. Expanding the Lagrangian in $\Phi$ then gives
\begin{equation}
\frac{\mathcal{L}_{\mathrm{NLSM}}}{\sqrt{-g}} =-\Tr\left[\tfrac{1}{2}\partial_{\mu}\Phi\partial^{\mu}\Phi+\Phi^{2}\partial_{\mu}\Phi\partial^{\mu}\Phi+\left(\Phi^{4}\partial_{\mu}\Phi\partial^{\mu}\Phi+\tfrac{1}{2}\Phi^{2}\partial_{\mu}\Phi\Phi^{2}\partial^{\mu}\Phi\right)+\mathcal{O}(\Phi^6)\right]. 
\label{NLSMexpand}
\end{equation}

The Lagrangians for the DBI and
sGal theory do not trivially lift to dS and were recently derived
from the following shift symmetry \cite{Bonifacio:2021mrf}:
\begin{equation}
\delta\phi=\theta_{A_{1}...A_{k}}X^{A_{1}}...X^{A_{k}}+...,
\end{equation}
where $X^{A}$ are embedding coordinates satisfying $-\left(X^{1}\right)^{2}+\sum_{A=2}^{d+2}\left(X^{A}\right)^{2}=1$ and the ellipsis denote field-dependent terms. This symmetry fixes the mass to be $m^{2}=-k(k+d)$. In the DBI case ($k=1$), the resulting
action is quite simple and given by
\eqs{
\frac{\mathcal{L}_{\mathrm{DBI}}}{\sqrt{-g}} = \frac{1}{(1-\phi^2)^\frac{d+1}{2}}\sqrt{1 - \frac{\nabla\phi\cdot\nabla\phi}{1-\phi^2}},
\label{dbilag}
}
where $\nabla\phi\cdot\nabla\phi=\partial_{\mu}\phi\partial^{\mu}\phi$. In the sGal case ($k=2$) the Lagrangian is very nontrivial:
\eqs{
\frac{\mathcal{L}_{\mathrm{sGal}}}{\sqrt{-g}}&= \bigg[\sum_{j=0}^{d}\frac{(1+\phi)^{d+1-j}+(-1)^j(1-\phi)^{d+1-j}}{2^{j+1}(1-\phi^2)^{\frac{d+4}{2}}\Gamma(j+3)}\left((j+1)f_{j+1}(\phi)-(j+2)f_j(\phi)\right)\partial^\mu\phi\partial^\nu\phi X^{(j)}_{\mu\nu}(\phi)\\
&\qquad - \frac{2}{d+2}\left(1 - \frac{(1+\phi)^{d+2}+(1-\phi)^{d+2}}{2(1-\phi^2)^{\frac{d+2}{2}}}\right)\bigg],
\label{sgalLag}
} 
where $X^{(j)}_{\mu\nu}$ is defined recursively as $X^{(n)}_{\mu\nu} = -n\nabla_\mu\nabla^\alpha\phi X^{(n{-}1)}_{\alpha\nu} + g_{\mu\nu}\nabla^\alpha\nabla^\beta\phi X^{(n{-}1)}_{\alpha\beta}$ with $X^{0}_{\mu\nu} = g_{\mu\nu}$, and 
\eq{
f_j(\phi) = {}_2F_1\left(\frac{d+4}{2},\frac{j+1}{2};\frac{j+3}{2};\frac{\nabla\phi\cdot\nabla\phi}{4(1-\phi^2)}\right) .
}

In the remainder of this paper, we will demonstrate that the masses and couplings of these
theories can be fixed by demanding that the wavefunction coefficients
have vanishing soft limits analogous to \eqref{enhanced}. 

\subsection{Wavefunction Coefficients}

We will work in the Poincare patch of dS$_{d+1}$ with metric
\begin{equation}
ds^{2}=\frac{1}{\eta^{2}}(d\vec{x}^{2}-d\eta^{2}), \label{eq:dSmetric}
\end{equation}
where $-\infty<\eta<0$ is the conformal time, and $\vec{x}$ denotes the boundary coordinate, with individual components $x^i$, $i=1,..,d$. Wavefunction coefficients $\Psi_n$ can be computed by analytic continuation of AdS Witten diagrams and thought of as $n$-point CFT correlators in the future boundary \cite{Maldacena:2002vr,McFadden:2009fg,McFadden:2010vh,Maldacena:2011nz,Raju:2011mp}. In momentum space, they can be expressed as
\begin{equation}
\Psi_{n}=\delta^{d}(\vec{k}_T)\langle\langle \mathcal{O}(\vec{k}_{1})...\mathcal{O}(\vec{k}_{n})\rangle\rangle ,
\label{psid}
\end{equation}
where $\vec{k}_T=\vec{k}_{1}+\ldots+\vec{k}_{n}$ is the sum of boundary momenta. The scalar operators $\mathcal{O}$ have scaling dimension $\Delta$, and are dual to scalar fields $\phi$ in the bulk with mass
\begin{equation}
m^{2}=\Delta(d-\Delta).
\label{eq:massdelta}
\end{equation}
In the previous subsection, we claimed that shift symmetries fix $m^2=-k(k+d)$ where $k=0,1,2$ for the NLSM, DBI, and sGal theories, respectively. The corresponding scaling dimensions are therefore $\Delta=d+k$. We will show that these values are required by enhanced soft limits of the wavefunction coefficients.

The bulk-to-boundary propagators in this background satisfy the free equations of motion $ (\mathcal{D}_k^{2}+m^{2})\phi^{\nu} =0$, where
\begin{equation}
\mathcal{D}_k^{2} = \eta^{2}\partial_{\eta}^{2}+(1-d)\eta\partial_{\eta}+\eta^{2}k^{2}, \label{eq:D2def}
\end{equation}
with $k=|\vec{k}|$. The solutions are given by
\begin{equation}
\phi^\nu(k, \eta) = (-1)^{\nu-\frac{1}{2}}\sqrt{\frac{\pi}{2}} k^\nu \eta^{d/2} H_\nu(-k \eta),
\label{bulktoboundaryprop}
\end{equation}
where $\nu=\Delta-d/2$, $H_\nu$ is a Hankel function of the second kind, and the normalisation is chosen for convenience. We then define an $n$-point contact diagram as follows:
\begin{equation}
\mathcal{C}^{\Delta}_n  =  \int  \frac{d\eta}{\eta^{d+1}} U_{1,n}(\eta),\,\,\,
U_{m,n}(\eta) =  \prod_{a=m}^{n}\phi_{a}, \label{eq:prodK}
\end{equation}
where $a$ labels an external leg, $k_a$ is the magnitude of the boundary momentum of that leg, and $\phi_a=\phi^\nu(k_{a},\eta)$.

As shown in \cite{Gomez:2021ujt}, soft limits of wavefunction coefficients take a particularly simple form when Witten diagrams are expressed in terms of certain differential operators constructed from boundary conformal generators acting on contact diagrams. The boundary conformal generators are given by  
\eqs{
P^i &= k^i,\\
D &= k^i\partial_i + (d-\Delta),\\
K_i &= k_i\partial^j\partial_j - 2k^j\partial_j\partial_i - 2(d-\Delta)\partial_i,\\
M_{ij} &= k_i\partial_j - k_j\partial_i,
\label{eqn:ConfGenerators}
}
where $\partial_i = \frac{\partial}{\partial k^i}$. We will collectively denote the generators by $\mathcal{D}^A \in\left\{ P^{i},M_{ij},D,K_{i}\right\} $, where $A$ is an adjoint index. Note that wavefunction coefficients satisfy the following conformal Ward identities:
\begin{equation}
\sum_{a=1}^{n}\mathcal{D}_{a}^{A}\Psi_{n}=0.
\end{equation}
Using boundary conformal generators we can define the following differential operators which will play an important role throughout the paper:
\eq{
\mathcal{D}_a\cdot\mathcal{D}_b = \frac{1}{2}\left(P^i_aK_{bi} + K_{ai}P^i_b - M_{a,ij}M_b^{ij}\right) + D_aD_b,
\label{casimir}
}
where $\mathcal{D}_a$ is a boundary conformal generator defined in terms of the boundary momentum associated with leg $a$. Acting on a pair of bulk-to-boundary propagators associated with legs $a$ and $b$, the operator in \eqref{casimir} satisfies the following useful identity:
\begin{equation}
(\mathcal{D}_{a}\cdot\mathcal{D}_{b})\left(\phi_{a}\phi_{b}\right)=\eta^{2}[\partial_{\eta}\phi_{a}\partial_{\eta}\phi_{b}+(\vec{k}_{a}\cdot\vec{k}_{b})\phi_{a}\phi_{b}].\label{eq:DaDb}
\end{equation}
Hence acting with $\mathcal{D}_{a}\cdot\mathcal{D}_{b}$ on a pair of bulk-to-boundary propagators is equivalent (up to a sign) to acting with a single $\nabla_a \cdot \nabla_b$, where $\nabla_a$ is a bulk covariant derivative acting on leg $a$. To simplify notation we will define $\hat{s}_{ab}=\mathcal{D}_{a}\cdot\mathcal{D}_{b}$.

In section \ref{6pt} we will also consider exchange diagrams so we need to define bulk-to-bulk propagators, $G_{\nu}(k,\eta,\tilde{\eta})$. For our purposes, we will only need to use the following property:
\begin{equation} \label{eq:exchnage=difxcontact}
[(\mathcal{D}_1+\ldots +\mathcal{D}_p)^2+m^2]^{-1} \mathcal{C}^{\Delta}_n =  \int  \frac{d\eta}{\eta^{d+1}}  \frac{d\tilde{\eta}}{\tilde{\eta}^{d+1}}   U_{p+1,n}(\eta) G_{\nu}(k_{1...p},\eta,\tilde{\eta})U_{1,p} (\tilde{\eta}).
\end{equation}
This follows from the equation of motion 
\begin{equation}\label{eq:eom-bulk-to-bulk}
(\mathcal{D}_{k}^{2}+m^{2})G_{\nu}=\eta^{d+1}\delta(\eta-\tilde{\eta}),
\end{equation}
and the following identity:
\begin{equation}
(\mathcal{D}_{1\ldots p}^{2}U_{1,p})U_{p{+}1, n}= (\mathcal{D}_{1}+\ldots + \mathcal{D}_{p})^2 U_{1,n}, \label{eq:bulkvsboundary-props}
\end{equation}
where in the left-hand side $\mathcal{D}_{1\ldots p}^{2}$ is defined in \eqref{eq:D2def} with $k=|\vec{k}_{1}+\ldots +\vec{k}_{p}|$ and $p<n$. For more details, see for example section 2.2 of \cite{Gomez:2021ujt}.

\section{Four-point soft limits} \label{fourpoint}

In this section, we will fix the masses and 4-point couplings of the NLSM, DBI, and sGal theories in de Sitter space from enhanced soft limits of their wavefunction coefficients. Our strategy will be to express the Witten diagrams in terms of differential operators acting on a contact diagram and then take the soft limit of a bulk-to-boundary propagator in the contact diagram. The soft limit of bulk-to-boundary propagators can be read off from the series expansion of \eqref{bulktoboundaryprop} which is schematically given by
\eq{
\phi^\nu(k, \eta) \sim \sum_{n=0}^\infty\left(a_{2n} + b_{2n} k^{2\Delta - d}\right)k^{2n}.
}
We can see that the second series has $k^{2(\nu+n)}$ terms which are subleading for positive $\nu$. In each case of interest, the enhanced soft limits will fix $\Delta = d + k$ where $k$ is the order of the shift symmetry in the Lagrangian. This sets $\nu=d/2+k$ and ensures that the second series does not contribute to the soft limit. We therefore take the soft limit of the wavefunction to be
\eqs{
\phi^\nu(k, \eta)&=\frac{\mathcal{N}}{\eta^{\Delta-d}}\left(1+\frac{\eta^2k^2}{2(2\Delta-d-2)}\right) + \mathcal{O}(k^4),\\
\mathrm{where}\,\, \mathcal{N} &=\frac{\Gamma\left(\Delta-d/2\right)2^{\Delta-d/2-1/2}}{\sqrt{\pi}}.
\label{eqn:SoftBtb}
}
This formula can then be used to study the soft limit of wavefunction coefficients. 

\subsection{NLSM}

The effective Lagrangian for the NLSM takes the following form at 4-points:
\begin{equation}
\frac{\mathcal{L}_{4}^{NLSM}}{\sqrt{-g}}= -\Tr\{\tfrac{1}{2}\nabla\Phi\cdot\nabla\Phi+\tfrac{1}{2}m^{2}\Phi^{2}+\Phi^{2}\nabla\Phi\cdot\nabla\Phi+\tfrac{1}{4}C\Phi^{4}\},
\label{nlsmlift}
\end{equation}
where we leave the mass and curvature correction $C$ unfixed. Note that the 2-derivative interaction comes from the naive uplift from flat space and we normalise the coupling to one. The corresponding tree-level flavour-ordered 4-point wavefunction coefficient can be obtained from two Witten diagrams and is given by \cite{Armstrong:2022csc}
\eqs{
\Psi_{4}^{NLSM}&=-\delta^{3}(\vec{k}_{T})\left(2\hat{s}_{13}+C-m^2\right)\mathcal{C}_{4}^{\Delta},\\ 
&=  -\delta^{3}(\vec{k}_{T})\int\frac{d\eta}{\eta^{d+1}}\left[2\eta^2\left(\vec{k}_1\cdot\vec{k}_3\phi_1\phi_3 + \dot{\phi}_1\dot{\phi}_3 \right)\phi_2\phi_4+ (C+\Delta(\Delta-d))\phi_1\phi_2\phi_3\phi_4\right].
\label{nlsmcorr}
}
If we take a soft limit of $\vec{k}_1$, we find that
\eqs{
\lim_{\vec{k}_{1}\rightarrow0}\Psi_{4}^{NLSM} &= \mathcal{N} \delta^{3}(\vec{k}_{T})\int\frac{d\eta}{\eta^{d+1}}\left[\eta^2\frac{2(\Delta-d)}{\eta^{\Delta-d+1}}\phi_2\dot{\phi}_3\phi_4 + \frac{C+\Delta(\Delta-d)}{\eta^{\Delta-d}}\phi_2\phi_3\phi_4\right] + \mathcal{O}(k_1),\\
& = \mathcal{N}\delta^{3}(\vec{k}_{T})\int\frac{d\eta}{\eta^{\Delta+1}}\left[2(\Delta-d)\eta\phi_2\dot{\phi}_3\phi_4 + \left(C+\Delta(\Delta-d)\right)\phi_2\phi_3\phi_4\right] + \mathcal{O}(k_1),\\
&= \mathcal{N}\delta^{3}(\vec{k}_{T})\left[2(\Delta-d)D_3+C-\Delta(\Delta-d)\right]\int\frac{d\eta}{\eta^{\Delta+1}}\phi_2\phi_3\phi_4 + \mathcal{O}(k_1),
\label{eq:NLSMSoft}
}
where in the final line we have used the definition of the dilatation operator acting on the bulk-to-boundary propagator. We see from \eqref{eq:NLSMSoft} that the soft limit will vanish to $\mathcal{O}(k_1)$ if $\Delta=d$ and $C=0$, i.e. if we have a massless scalar and no curvature corrections in agreement with \eqref{NLSMexpand}. We can also see from \eqref{nlsmcorr} that it is not possible to for the soft limit to vanish at higher order since there is no way to cancel the $\vec{k}_1 \cdot \vec{k}_3$ term given that the bulk-to-boundary propagators only depend on magnitudes of momenta. Hence, the wavefunction coefficient is simply 
\begin{equation}
\Psi_{4}^{NLSM}=-2\delta^{3}(\vec{k}_{T})\hat{s}_{13}\mathcal{C}_{4}^{\Delta=d}.
\label{4ptnlsmwavefunctionfinal}
\end{equation} 

\subsection{DBI}

At 4-points, the DBI theory can be described by the following general effective Lagrangian (modulo integration by parts and free equations of motion):
\begin{equation}
\frac{\mathcal{L}^{DBI}_4}{\sqrt{-g}}= -\{\tfrac{1}{2}\nabla\phi\cdot\nabla\phi+\tfrac{1}{2}m^{2}\phi^{2}+\tfrac{1}{8}(\nabla\phi\cdot\nabla\phi)^{2}+\tfrac{1}{4!}C\phi^{4}\},\label{effectiveaction}
\end{equation}
where the 4-derivative interaction (whose coupling we have normalised to one) arises from the naive uplift from flat space and we leave the mass and curvature correction $C$ unfixed. The tree-level 4-point wavefunction coefficient can be computed from Witten diagrams and is given by \cite{Armstrong:2022csc}
\eq{
\Psi_4^{DBI} = -\delta^{3}\left(\vec{k}_{T}\right)\left(\hat{s}_{12}^2 + \hat{s}_{13}^2 + \hat{s}_{14}^2 + C\right)\mathcal{C}^{\Delta}_4.
\label{dbi4ptwavfnct}
}
More explicitly, the action of $\hat{s}_{12}^2$ on bulk-to-boundary propagators is given by 
\eqs{
\hat{s}_{12}^2 \phi_1\phi_2 &= \eta^4\Big[(\vec{k}_1\cdot\vec{k}_2)^2\phi_1\phi_2 + 2\vec{k}_1\cdot\vec{k}_2\dot{\phi}_1\dot{\phi}_2 + \ddot{\phi}_1\ddot{\phi}_2,\\
&\qquad + \frac{1}{\eta}\left(2\vec{k}_1\cdot\vec{k}_2\left(\phi_1\dot{\phi}_2 + \dot{\phi}_1\phi_2\right) -k_1^2\phi_1\dot{\phi}_2 -k_2^2\dot{\phi}_1\phi_2 + \dot{\phi}_1\ddot{\phi}_2 + \ddot{\phi}_1\dot{\phi}_2\right)\\
&\qquad + \frac{1}{\eta^2}\left((2-d)\vec{k}_1\cdot\vec{k}_2\phi_1\phi_2 + \dot{\phi}_1\dot{\phi}_2\right)\Big].
 \label{eq:DBIIntegrand}
}
We then insert the soft limit for $\phi_1$ from equation (\ref{eqn:SoftBtb}). 

To fix $\Delta$ and $C$ we need to expand the integrand to $\mathcal{O}\left(k_{1}^{2}\right)$ and use the equations of motion and integration by parts to eliminate terms which are not independent. One option is to use the equations of motion of the bulk-to-boundary propagators to remove any explicit dependence on $k_2^2$ in \eqref{eq:DBIIntegrand} ($k_2$ will still enter in the arguments of $\phi_2$). Alternatively, we can apply the equations of motion to leave only terms containing $\phi_2$ and $\dot{\phi}_2$ along with factors of $k_2^2$. This second approach is equivalent to using the identity $H_{\nu-1}(x)=-H_{\nu+1}(x) + \frac{2\nu}{x}H_\nu(x)$ on the Hankel functions which appear in the derivatives of propagators to leave only two independent functions. Removing the explicit dependence on $k_2^2$ in the first term of \eqref{dbi4ptwavfnct} and summing over cyclic permutations then gives
\eqs{
\lim_{\vec{k}_{1}\rightarrow0}\Psi_{4}^{DBI}&= \mathcal{N}\delta^{3}\left(\vec{k}_{T}\right)\int\frac{d\eta}{\eta^{\Delta+1}}\bigg[(\Delta-d-1)\left((\Delta-d)\eta^2\ddot{\phi}_2 - 2\eta^3\vec{k}_1\cdot\vec{k}_2\dot{\phi}_2\right) \phi_3\phi_4\\
&\qquad+ \mathrm{Cyc.}[234] + \left(\Delta(\Delta-d)(4\Delta-3d-1)+C\right)\phi_2\phi_3\phi_4 + \mathcal{O}(k_1^2)\bigg],
\label{eqn:4ptDBISoft}
}
where we used the following identity to remove the $\dot{\phi}_a$ terms ($a \in\left\{ 2,3,4\right\} $) at $\mathcal{O}(k_1^0)$:
\eq{
\int\frac{d\eta}{\eta^{\Delta+1}}\eta\partial_\eta\left(\prod_{i=2}^n\phi_i\right) \sim \Delta \int\frac{d\eta}{\eta^{\Delta+1}}\left(\prod_{i=2}^n\phi_i\right).
\label{eqn:IntegrandIBP}
}
In deriving the above formula, we discarded a total derivative term. This term actually gives divergent contributions at $\eta=0$ and therefore needs to be regulated, however these contributions are analytic in at least two momenta and therefore correspond to contact terms which have delta function support when Fourier transformed to position space \cite{Maldacena:2011nz}. 

From \eqref{eqn:4ptDBISoft}, we see that the soft limit vanishes to $\mathcal{O}(k_1^2)$ if $\Delta=d+1$ and $C = -(d+1)(d+3)$. Plugging these values into \eqref{dbi4ptwavfnct}  gives
\begin{equation}
\Psi_{4}^{DBI}=-\delta^{3}\left(\vec{k}_{T}\right)\left(\hat{s}_{12}^{2}+\hat{s}_{13}^{2}+\hat{s}_{14}^{2}-(d+1)(d+3)\right)\mathcal{C}_{4}^{\Delta=d+1}.
\label{dbi4ptfinal}
\end{equation}
Moreover, \eqref{effectiveaction} becomes 
\begin{equation}
\frac{\mathcal{L}_{4}^{DBI}}{\sqrt{-g}}=-\{\tfrac{1}{2}\nabla\phi\cdot\nabla\phi-\tfrac{d+1}{2}\phi^{2}+\tfrac{1}{8}(\nabla\phi\cdot\nabla\phi)^{2}-\tfrac{(d+1)(d+3)}{4!}\phi^{4}\}.
\label{dbi4pt}
\end{equation}
From \eqref{eq:DBIIntegrand} we can see that it is not possible for the soft limit to vanish beyond $\mathcal{O}(k_1^2)$ since this term contains a piece proportional to $(\vec{k}_1\cdot\vec{k}_2)^2$ but the soft limit of Witten diagrams coming from the $\phi^4$ interaction will only depend on the magnitude $k_1$. We also note that while the $\mathcal{O}(k_1)$ contribution to the wavefunction coefficient is needed to fix $\Delta$, once this is fixed only the leading soft limit is needed to fix $C$. This is appears to be a general feature in de Sitter space, in contrast to flat space where all the subleading data is needed to fix coefficients. 

Let us now compare to the Lagrangian in \eqref{dbilag} which was derived from shift symmetries. Expanding it to quartic order gives 
\eqs{
\frac{\mathcal{L}_{\mathrm{DBI}}}{\sqrt{-g}} &= \frac{1}{(1-\phi^{2})^{(d+1)/2}}\sqrt{1-\frac{\nabla\phi\cdot\nabla\phi}{1-\phi^{2}}},\\
&= -\left(\frac{1}{2}\nabla\phi\cdot\nabla\phi\ensuremath{-\frac{d+1}{2}\phi^{2}+\frac{1}{8}(\nabla\phi\cdot\nabla\phi)^{2}-\frac{(d+1)(d+3)}{4!}\phi^{4}+\mathcal{O}(\phi^{6})}\right),
}
where we have used integration by parts and the free equation of motion $\nabla^2\phi = m^2\phi$ to remove a $(\nabla\phi\cdot\nabla\phi)\phi^2$ term. This precisely matches \eqref{dbi4pt}, which was derived from enhanced soft limits.

\subsection{sGal} \label{sgal4pt}

At 4-points, the sGal theory can be described by the following effective action modulo integration by parts and free equations of motion:
\begin{equation}
\frac{\mathcal{L}^{sGal}_4}{\sqrt{-g}}= -\{\tfrac{1}{2}\nabla\phi\cdot\nabla\phi+\tfrac{1}{2}m^{2}\phi^{2}+\tfrac{1}{8}(\nabla_{\mu}\nabla_{\nu}\phi)^{2}\nabla\phi\cdot\nabla\phi+\tfrac{1}{8}B(\nabla\phi\cdot\nabla\phi)^{2}+\tfrac{1}{4!}C\phi^{4}\},\label{effectiveaction6derv}
\end{equation}
where the 6-derivative term uplifts from flat space and we have normalised its coupling to one while the remaining interaction terms are curvature corrections with unfixed coefficients.  The 4-point wavefunction coefficient can be computed from Witten diagrams and is given by \cite{Armstrong:2022csc}
\begin{equation}
\Psi^{sGal}_{4}=\delta^{3}(\vec{k}_{T})[(\hat{s}_{12}^{3}+\hat{s}_{13}^{3}+\hat{s}_{14}^{3})+(d-B)(\hat{s}_{12}^{2}+\hat{s}_{13}^{2}+\hat{s}_{14}^{2})-C]\mathcal{C}_{4}^{\Delta}.
\label{4pteffectivesgal}
\end{equation}
The $\hat{s}_{ab}^3$ terms are quite lengthy and can be found in Appendix \ref{app:sGal}. The $\hat{s}_{ab}^2$ terms were already considered in the previous subsection. 

We will now expand the integrand up to $\mathcal{O}(k_1^2)$ and present the soft limit in parts. After substituting (\ref{eqn:SoftBtb}) we apply equations of motion to eliminate any explicit dependence on $k_2^2$ in the $\hat{s}_{12}^{3}$ term and sum over permutations to obtain
\eqs{
\lim_{\vec{k}_{1}\rightarrow0}\Psi_{4}^{sGal} &= -\mathcal{N}(\Delta-d-2)\delta^{3}\left(\vec{k}_{T}\right)\int\frac{d\eta}{\eta^{\Delta+1}}\bigg[\eta\bigg((\Delta-d-1)(\Delta-d)\eta^2\\
&\qquad\qquad\qquad\qquad  + \frac{k_1^2}{2\Delta-d-2}(\Delta-d-3)(\Delta-d-4)\bigg)\dddot{\phi}_2\\
&\qquad\qquad - 3\vec{k}_1\cdot\vec{k}_2\eta^4\ddot{\phi}_2 + 3(\vec{k}_1\cdot\vec{k}_2)^2\eta^5\dot{\phi}_2\bigg]\phi_3\phi_4 + \mathrm{Cyc.}[234] +  \mathcal{O}(k_1^3) + \dots,
}
where the ellipsis represent terms that can also arise from 4-derivative and $\phi^4$ interactions. We must then set $\Delta = d+2$ in order for the terms displayed above to vanish. When this is substituted into the remaining terms they simplify significantly and we obtain
\eqs{
\lim_{\vec{k}_{1}\rightarrow0}\Psi_{4}^{sGal}&= -\mathcal{N}(B+2d+2)\delta^{3}\left(\vec{k}_{T}\right)\int\frac{d\eta}{\eta^{\Delta+1}}\eta^2\left(2\ddot{\phi}_2 - 2\eta(\vec{k}_1\cdot\vec{k}_2)\dot{\phi}_2 + \eta^2(\vec{k}_1\cdot\vec{k}_2)^2\phi_2\right)\phi_3\phi_4\\
&\qquad\qquad\qquad\qquad\qquad\qquad + \mathrm{Cyc.}[234] +  \mathcal{O}(k_1^3) + \dots,
\label{eqn:sGalsoftB}
}
where the ellipsis denote terms that can also arise from $\phi^4$ interactions. After setting $B=-2(d+1)$ the above terms vanish and the soft limit of the wavefunction coefficient reduces to
\eq{
\lim_{\vec{k}_{1}\rightarrow0}\Psi_{4}^{sGal} = \mathcal{N}(4(d+2)^2-C)\delta^{3}\left(\vec{k}_{T}\right)\int\frac{d\eta}{\eta^{\Delta+1}}\frac{4+2d+\eta^2k_1^2}{2(d+2)}\phi_2\phi_3\phi_4 +  \mathcal{O}(k_1^3),
\label{eqn:sGalsoftC}
}
which fixes $C = 4(d+2)^3$. The wavefunction coefficient with $\mathcal{O}(k_1^3)$ soft behavior is therefore 
\eq{
\Psi_{4}^{sGal} = \delta^{3}\left(\vec{k}_{T}\right)\left(\hat{s}_{12}^{3}+\hat{s}_{13}^{3}+\hat{s}_{14}^{3}+(3d+2)\left(\hat{s}_{12}^{2}+\hat{s}_{13}^{2}+\hat{s}_{14}^{2}\right)-4(d+2)^{3}\right)\mathcal{C}^{\Delta=d+2}_{4}.
\label{sGalwavefunction4pt}
}
We can see from equations (\ref{eqn:sGalsoftB}) and (\ref{eqn:sGalsoftC}) that once $\Delta$ is fixed, we can fix $B$ and $C$ using only the leading order soft limit.

Moreover, we find that the Lagrangian in \eqref{effectiveaction6derv} is given by
\begin{equation}
\frac{\mathcal{L}_{4}^{sGal}}{\sqrt{-g}}=-\{\tfrac{1}{2}\nabla\phi\cdot\nabla\phi-(d+2)\phi^{2}+\tfrac{1}{8}(\nabla_{\mu}\nabla_{\nu}\phi)^{2}\nabla\phi\cdot\nabla\phi-\tfrac{d+1}{4}(\nabla\phi\cdot\nabla\phi)^{2}+\tfrac{(d+2)^{3}}{6}\phi^{4}\}.
\label{effectivelagrangiansgalenhanced}
\end{equation}

Let us compare the above Lagrangian to the one derived from hidden symmetry. Expanding \eqref{sgalLag} to quartic order gives
\eqs{
\frac{\mathcal{L}^{\mathrm{sGal}}}{\sqrt{-g}}&= -\bigg(\frac{1}{2}\nabla\phi\cdot\nabla\phi-(d+2)\phi^{2}-\frac{1}{4!}2(d+2)(d(d+4)+12)\phi^{4}+\frac{1}{4!}(d(3d+8)+28)\phi^{2}\nabla\phi\cdot\nabla\phi\\
&\qquad\qquad+\frac{d+4}{96}(\nabla\phi\cdot\nabla\phi)^{2}+ \frac{2-d}{24}\phi\nabla^\mu\phi\nabla^\nu\phi\nabla_\mu\nabla_\nu\phi - \frac{1}{96}\nabla\phi\cdot\nabla\phi\left(\nabla_{\mu}\nabla_{\nu}\phi\right)^{2}\\&\qquad\qquad  + \frac{1}{48}\nabla^\mu\phi\nabla^\nu\phi\nabla_\sigma\nabla_\mu\phi\nabla^\sigma\nabla_\nu\phi\bigg)+\mathcal{O}(\phi^6),
\label{eqn:sGal4ptLag}
}
where we have used the free equation of motion $\nabla^2\phi = m^2\phi = -2(d+2)\phi$. We can then use integration by parts and free equations of motion to bring this to the form in \eqref{effectiveaction6derv}. In more detail,  the final term in (\ref{eqn:sGal4ptLag}) can be written as
\eq{
\partial^\mu\phi\partial^\nu\phi\nabla^\sigma\nabla_\mu\phi\nabla_\sigma\nabla_\nu\phi \sim -\frac{1}{2}\left((\nabla\phi\cdot\nabla\phi)\nabla^\sigma\nabla^\mu\phi\nabla_\sigma\nabla_\mu\phi + (\nabla\phi\cdot\nabla\phi)\partial^\nu \phi \nabla^2\nabla_\nu\phi\right),
\label{simplify1}
}
where we applied integration by parts on $\nabla^{\sigma}$. The second term on the right-hand side can then be reduced to lower-derivative terms by noting that
\eqs{
\nabla_\sigma\nabla^\sigma\partial_\nu\sigma\phi &= \nabla_\sigma\nabla_\nu\partial^\sigma\phi,\\
&=\nabla_\nu\nabla^2\phi + [\nabla_\nu\nabla_\sigma]\partial^\sigma\phi,\\
&= m^2\partial_\nu\phi + R_{\mu\nu}\partial^\mu\phi,\\
&= -(d+4)\partial_\nu\phi.
}
Similarly, using integration by parts and free equations of motion, the two-derivative term in the first line of \eqref{eqn:sGal4ptLag} can be reduced to a $\phi^4$ term, and the second four-derivative term in the second line of \eqref{eqn:sGal4ptLag} can be written in the form $(\nabla\phi\cdot\nabla\phi)^{2}$ plus a $\phi^4$ term. In the end, we are left with three interaction terms:
\eq{
\frac{\mathcal{L}^{\rm{sGal}}_{\rm{int}}}{\sqrt{-g}} = -\frac{1}{48}(\nabla\phi\cdot\nabla\phi)\nabla^\alpha\nabla^\beta\phi\nabla_\alpha\nabla_\beta\phi + \frac{d+1}{24}(\nabla\phi\cdot\nabla\phi)^{2} - \frac{1}{36}(d+2)^3\phi^4+\mathcal{O}(\phi^6).
}
After multiplying by 6 (equivalent to rescaling the six-derivative coupling) this indeed matches the interaction terms in \eqref{effectivelagrangiansgalenhanced}, which were deduced from enhanced soft limits. 

\subsection{Double Copy} \label{doublec} 

In flat space, the scattering amplitudes of the NLSM, DBI, and sGal theories enjoy double copy relations \cite{Cachazo:2014xea}, which are made manifest using a formulation based on scattering equations \cite{Cachazo:2013hca,Mason:2013sva}. Scattering equations in (A)dS were later formulated in \cite{Eberhardt:2020ewh,Roehrig:2020kck,Gomez:2021qfd,Gomez:2021ujt} and used to explore the double copy for effective scalar theories in \cite{Armstrong:2022csc} (the double copy for effective scalar theories in AdS was also explored from various other points of view in \cite{Sivaramakrishnan:2021srm,Herderschee:2022ntr,Cheung:2022pdk}). In more detail, a generalised double copy for 4-point wavefunction coefficients was proposed in terms of unfixed parameters encoding masses and curvature corrections. In this subsection, we will explain how to fix these parameters using our results on enhanced soft limits. 

Let us briefly review the representation of tree-level wavefunction coefficients in terms of scattering equations and the generalised double copy at 4-points. We will focus on effective scalar theories with mass $m^2=\Delta(d-\Delta)$. The discussion will be very schematic but the interested reader can find more details in \cite{Armstrong:2022csc}. A tree-level $n$-point wavefunction coefficient can be written as an integral over $n$-punctures on the sphere: 
\begin{equation}
\Psi_{n}=\delta^{d}(\vec{k}_{T})\int_{\gamma}\prod_{a\neq e,f,g}^{n}\dif\sigma_{a}\,S_{a}^{-1}(\s_{ef}\s_{fg}\s_{ge})^{2}\,\,{\cal I}_{n}{\cal C}_{n}^{\Delta},
\label{worldsheetintegral}
\end{equation}
where $\sigma_{ab}=\sigma_{a}-\sigma_{b}$. The three punctures denoted $e,f,g$ are fixed and ${\cal I}_{n}$ is a theory-dependent integrand, which in general is a differential operator acting on an $n$-point contact diagram $\mathcal{C}_n^{\Delta}$. Since the integrand is constructed from $\hat{s}_{ab}$ operators it can in principle have ordering ambiguities, although they do not arise for scalar theories with polynomial interactions \cite{Gomez:2021qfd}. The contour $\gamma$ encircles the poles where differential operators $S_a$ vanish when acting on everything to the right. The operators are defined as
\begin{equation}
S_{a}=\sum_{{j=1\atop j\neq i}}^{n}\frac{\alpha_{ab}}{\sigma_{ab}}
\end{equation}
where $\alpha_{ab}=2\hat{s}_{ab}+\mu_{ab}$ with $\mu_{aa\pm1}=-m^{2}$. In practice, it is not known how to explicitly solve the equations which determine these poles, dubbed the cosmological scattering equations, but the integral can be mapped to a sum of Witten diagrams using the global residue theorem. 

For the NLSM at 4-points, the following integrand was proposed in \cite{Armstrong:2022csc}:
\begin{equation}
\mathcal{I}_{4}^{NLSM}=\lambda^{2}{\rm PT}\left({\rm Pf}'A\right)^{2}+c{\rm PT}\left.{\rm Pf}X\right|_{\rm{conn}}{\rm Pf'}A,
\label{nlsmintegrand}
\end{equation}
where ${\rm PT}=\left(\sigma_{12}...\sigma_{n1}\right)^{-1}$, ${\rm Pf'}A$ is related to the Pfaffian of an operator-valued matrix whose off-diagonal elements are $A_{rs}=\a_{rs}/\sigma_{rs}$, ${\rm Pf}X$ is the Pfaffian of a matrix whose off-diagonal elements are $X_{rs}=1/\sigma_{rs}$, and $\left.{\rm Pf}\,X\right|_{\rm{conn}}$ refers to the sum over connected perfect matchings which arise in ${\rm Pf}X$. The first term on the right-hand side of \eqref{nlsmintegrand} represents the naive uplift from flat space while the second term encodes a potential curvature correction. Evaluating the contour integral in \eqref{worldsheetintegral} then gives 
\begin{equation}
\Psi_{4}^{NLSM}=-\delta^{3}(\vec{k}_{T})\left(2\lambda^{2}\hat{s}_{13}-c-m^{2}\right)\mathcal{C}_{4}^{\Delta}.
\end{equation}
Comparing this to the wavefunction coefficient with enhanced soft limits in \eqref{4ptnlsmwavefunctionfinal} then fixes the mass and coefficients as follows:
\begin{equation}
\lambda=1,\,\,\,c=m=0.
\end{equation}

For the DBI and sGal theories at four-points the following integrand was proposed in \cite{Armstrong:2022csc}: 
\begin{multline}
\mathcal{I}_{4}^{(6)}=a({\rm Pf}'A)^{3}({\rm Pf'}A+m^{2}\left.{\rm Pf}X\right|_{{\rm conn}}) +b({\rm Pf}'A)^{2}({\rm Pf'}A{\rm Pf}X+m^{2}{\rm PT})+c {\rm PT}\left.{\rm Pf}X\right|_{\rm{conn}}{\rm Pf'}A,
\label{6derivintegrand}
\end{multline}
where $a=0$ for the DBI theory (note that in the above equation $a,b,c$ are understood to be coefficients rather than labels of external legs). For both theories, $c$ is a curvature correction while $b$ is also a curvature correction in the sGal theory. Note that \eqref{6derivintegrand} can be obtained from \eqref{nlsmintegrand} via the following replacement: 
\begin{equation}
\lambda^2 {\rm PT}\rightarrow a{\rm Pf'}A\left({\rm Pf'}A+m^{2}\left.{\rm Pf}X\right|_{{\rm conn}}\right)+b\left({\rm Pf'}A{\rm Pf}X+m^{2}{\rm PT}\right).
\end{equation}
In addition to performing this replacement, we are also free to change the value of the mass and coefficient $c$ in \eqref{nlsmintegrand} so that they do not necessarily have the same value as the NLSM. In the flat space limit (where curvature corrections and masses are set to zero), this replacement encodes the double copy of NLSM amplitudes to DBI and sGal amplitudes. In curved background, we therefore refer to it as a generalised double copy. 

After specifying a simple prescription to avoid potential ordering ambiguities of the integrand in \eqref{6derivintegrand}, the contour integral in \eqref{worldsheetintegral} gives
\begin{equation}
\Psi^{(6)}_{4}=\delta^{3}(\vec{k}_{T})[\frac{8a}{3}(\hat{s}_{12}^{3}+\hat{s}_{13}^{3}+\hat{s}_{14}^{3})+2(b-am^{2})(\hat{s}_{12}^{2}+\hat{s}_{13}^{2}+\hat{s}_{14}^{2})+\frac{1}{3}am^{6}-bm^{4}+c]\mathcal{C}_{4}^{\Delta}.
\label{general4ptwavefuntion}
\end{equation}
Comparing this to the wavefunction coefficient for the DBI theory derived from enhanced soft limits in \eqref{dbi4ptfinal} then fixes the parameters as follows:
\begin{equation}
a=0,\,\,\,b=-\frac{1}{2},\,\,\,c=\frac{1}{2}\left(d^{2}+6d+5\right),\,\,\,m^{2}=-(d+1).
\end{equation}
Moreover, comparing \eqref{general4ptwavefuntion} to the wavefunction coefficient for the sGal theory in \eqref{sGalwavefunction4pt} implies that
\begin{equation}
a=\frac{3}{8},\,\,\,b=\frac{1}{4}\left(3d-2\right),\,\,\,c=-8(d+2)^{2},\,\,\,m^{2}=-2(d+2).
\end{equation}

In summary, the parameters of the generalised double copy for four-point wavefunction coefficients can be fully fixed by enhanced soft limits. In the next section we will show that enhanced soft limits also fix higher-point wavefunction coefficients, so it would be interesting to see if the double copy prescription can be extended to higher points as well.

\section{Higher Points} \label{6pt}

In this section, we will show that all 6-point couplings of the NLSM and DBI theory in dS can also be fixed from enhanced soft limits of wavefunction coefficients. The method we develop can also be applied to the sGal theory, but at six points its Lagrangian has 13 interaction vertices going up to ten derivatives so Witten diagram calculations become very tedious. We will therefore leave that case for future work. 

\subsection{NLSM}

We start with the NLSM, which is very simple but nicely illustrates the procedure for fixing higher-point couplings. At six points, the most general Lagrangian is given by
\eqs{
\frac{\mathcal{L}^{NLSM}_{6}}{\sqrt{-g}}={\rm Tr}\left[-\frac{1}{2}\partial_\mu \Phi \partial^\mu \Phi-\frac{1}{2}m^{2}\Phi^{2}-\Phi^{2}\partial_{\mu}\Phi\partial^{\mu}\Phi-\frac{1}{4}C\Phi^{4}\right. \\
\quad \left.-A\left(\Phi^{4}\partial_{\mu}\Phi\partial^{\mu}\Phi+\frac{1}{2}\Phi^{2}\partial_{\mu}\Phi\Phi^{2}\partial^{\mu}\Phi\right)-\frac{1}{6}F\Phi^{6}\right],
\label{6ptnlsmlagrangian}
}
where the $\Phi^4$ and $\Phi^6$ terms are curvature corrections. We have already fixed $m=0$ and $C=0$ from the enhanced soft limit at four points. The coefficient $A$ can be fixed by the flat space limit but we will deduce it along with $F$ from enhanced soft limits at six points. The six-point wavefunction coefficient was already computed from Witten diagrams in \cite{Armstrong:2022csc} and takes the form 
\eq{
\Psi^{NLSM}_6 = \delta^{3}(\vec{k}_{T}) \left[\left(\frac{\hat{s}_{13}\hat{s}_{46}}{\hat{s}_{123}} + A\,\hat{s}_{13} + \mathrm{Cyc}.[i\to i{+}2]\right) + F\right]\mathcal{C}^{\Delta=d}_6,
\label{nlsm6pt}
}
where we've used the shorthand and $\hat{s}_{abc} =\mathcal{D}_a\cdot\mathcal{D}_b+\mathcal{D}_b\cdot\mathcal{D}_c+\mathcal{D}_c\cdot\mathcal{D}_a$. The first term in parenthesis comes from an exchange diagram with two 4-point vertices. It was obtained using integration by parts to move all derivatives with respect to conformal time onto the external propagators. In this form, the expression is free or ordering ambiguities since $[\hat{s}_{abc},\hat{s}_{ab}]\mathcal{C}^\Delta = 0$.

If we take $\vec{k}_1$ soft, all operators of the form $\mathcal{D}_1\cdot\mathcal{D}_a$ will vanish up to $\mathcal{O}(k_1)$ when acting on the contact diagram $\mathcal{C}^{\Delta}$ as in (\ref{eq:NLSMSoft}) since $\Delta=d$. Hence two of the channels in \eqref{nlsm6pt} drop out immediately and it reduces to
\eq{
\lim_{\vec{k}_{1}\to0}\Psi_{6}^{NLSM}= \delta^{3}(\vec{k}_{T})\left[\left(\frac{\hat{s}_{35}\hat{s}_{62}}{\hat{s}_{612}} + A\,\hat{s}_{35}\right) + F\right]\mathcal{C}^{\Delta=d}_6.
}
Noting that $\lim_{\vec{k}_1\to 0}\hat{s}_{612} = \hat{s}_{62}$ when $\Delta=d$, we then can see the soft limit vanishes if $A=-1$ and $F=0$, in agreement with \eqref{NLSMexpand}.

In summary, we see that the enhanced soft limit arises via cancellations between exchange and contact diagrams, fixing higher-point couplings in terms of lower-point couplings. In this way, we can in principle bootstrap all tree-level wavefunction coefficients and reconstruct the Lagrangian.

\subsection{DBI} \label{dbi6point} 

\begin{figure}
\centering
\includegraphics[width=12cm]{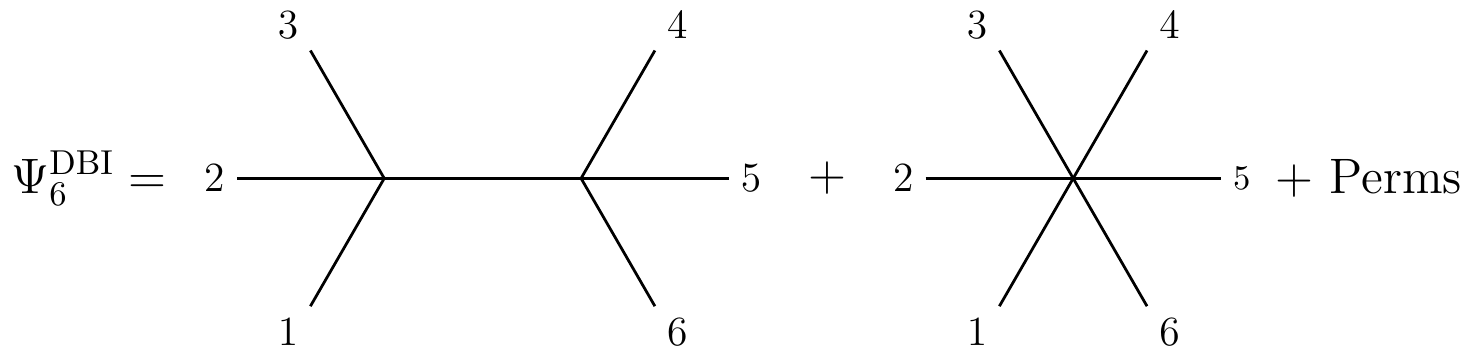}
\caption{Witten diagrams contributing the 6-point sGal wavefunction coefficient.}
\label{fig:6ptDiagram}
\end{figure}

We now consider the following 6-point effective Lagrangian:
\begin{equation}
\frac{\mathcal{L}_{6}^{DBI}}{\sqrt{-g}}=\frac{\mathcal{L}_{4}^{DBI}}{\sqrt{-g}}+\frac{A}{48}(\nabla\phi\cdot\nabla\phi)^{3}+\frac{B}{16}(\nabla\phi\cdot\nabla\phi)^{2}\phi^{2}+\frac{C}{6!}\phi^{6},
\label{dbilagrangian}
\end{equation}
where the 4-point Lagrangian was fixed by enhanced soft limits in \eqref{dbi4pt}. The coefficient $A$ can be determined by the flat space limit but we will fix it along with the other coefficients from enhanced soft limits. First we compute the 6-point wavefunction coefficient from Witten diagrams, which are depicted in Figure \ref{fig:6ptDiagram}.

\begin{figure}
\centering
\includegraphics[width=4.6cm]{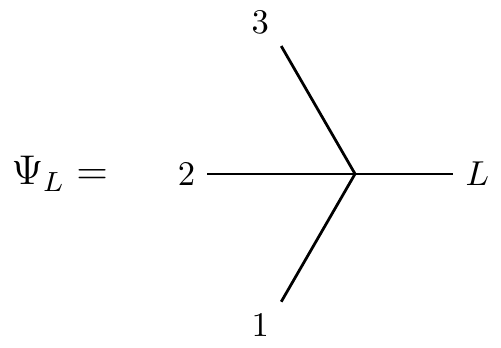}
\caption{Four-point vertex contributing to 6-point exchange diagram.}
\label{fig:6ptIBP}
\end{figure}

To compute the exchange diagrams, first consider the 4-point vertex on the left of the exchange diagram in Figure \ref{fig:6ptDiagram} which is illustrated in Figure \ref{fig:6ptIBP}:
\eq{
\Psi_L = \left(\hat{s}_{12}\hat{s}_{3L} + \hat{s}_{23}\hat{s}_{1L} + \hat{s}_{31}\hat{s}_{2L} - (d+1)(d+3)\right),
\label{eqn:DBIHalfExch}
}
which is understood to act on a 6-point contact diagram in combination with a bulk-to-bulk propagator and another 4-point vertex. 
We can then use the conformal Ward identities at the vertex $\mathcal{D}_1+\mathcal{D}_2+\mathcal{D}_3=-\mathcal{D}_L$ to get
\eq{
\Psi_L = \big(-2\left(\hat{s}_{12}\hat{s}_{23} + \hat{s}_{23}\hat{s}_{31} + \hat{s}_{31}\hat{s}_{12}\right) + m^2(\hat{s}_{12}+\hat{s}_{23}+\hat{s}_{31}) - (d+1)(d+3)\big),
\label{4ptL}
}
where $-m^2=\Delta(\Delta-d)=d+1$. Combining this with the rest of the Witten diagram and summing over permutations then gives
\eqs{
\Psi^{DBI}_{6,\, \mathrm{exch}} &= \delta^{3}(\vec{k}_{T}) \frac{1}{ \left(\mathcal{D}_1+\mathcal{D}_2+\mathcal{D}_3\right)^2+m^2}\Big(2\left(\hat{s}_{12}\hat{s}_{23}+\hat{s}_{23}\hat{s}_{31}+\hat{s}_{31}\hat{s}_{12}\right)\\
&\qquad\qquad + (d+1)(\hat{s}_{12}+\hat{s}_{23}+\hat{s}_{31}+(d+3))\Big)\times(123)\to(456))\mathcal{C}_6^{\Delta=d+1} + \mathrm{perms},
\label{dbiexchange}
}
where the permutation sum is over 10 inequivalent factorisation channels. Note that this expression is free of ordering ambiguities. Moreover, it is straightforward to read off the contact Witten diagrams from \eqref{dbilagrangian}:
\eq{
\Psi^{DBI}_{6,\, \mathrm{cont}} = \delta^{3}(\vec{k}_{T}) \left[A\left(\hat{s}_{12}\hat{s}_{34}\hat{s}_{56} + \mathrm{perms}\right) + B\left(\hat{s}_{12}\hat{s}_{34} + \mathrm{perms}\right) + C\right]\mathcal{C}^{\Delta=d+1}_6,
}
where we sum over inequivalent permutations giving 61 terms.

Let us now expand the wavefunction coefficient to $\mathcal{O}(k_1^2)$. To this order, the 4-point vertex in \eqref{4ptL} is given by 
\eq{
\Psi_L = -( \left(\mathcal{D}_1+\mathcal{D}_2+\mathcal{D}_3\right)^2+m^2)\left(\hat{s}_{12}+\hat{s}_{31}+\frac{1}{2}(d-1)\right) + \mathcal{O}(k_1^2),
\label{DBI6ptExchLim}
}
As a result, the numerator of the exchange diagram in Figure \ref{fig:6ptDiagram} can be written as 
\eqs{
\Big(2\left(\hat{s}_{12}\hat{s}_{23}+\hat{s}_{23}\hat{s}_{31}+\hat{s}_{31}\hat{s}_{12}\right) + (d+1)(\hat{s}_{12}+\hat{s}_{23}+\hat{s}_{31}+(d+3))\Big) \\
= ( \left(\mathcal{D}_1+\mathcal{D}_2+\mathcal{D}_3\right)^2+m^2)\left(\hat{s}_{12}+\hat{s}_{31} + \frac{1}{2}(d-1)\right)+\mathcal{O}(k_1^2).
}
Hence, in the soft limit we can cancel all the propagators and are left with a cubic polynomial in $\hat{s}_{ij}$. We then we apply conformal Ward identities to cancel exchange and contact contributions, mimicking the analogous cancellation of terms that arises in the flat space limit using momentum conservation.

We then use integration by parts and equations of motion to write the conformal time integrand in terms of linearly independent terms, as before. In the present case, the procedure is somewhat complicated so we provide more details in Appendix \ref{6ptdbiappendix} and the attached Mathematica code \verb|EnhancedSoftLimits.nb|. In the end, we find that the soft limit of the 6-point wavefunction coefficient vanishes to $\mathcal{O}(k_1^2)$ if and only if $A=3, B=d+1,C=2(d+1)(9-d^2)$. Since $\Delta$ was already fixed from the 4-point soft limit, these values can be deduced by considering only the leading order soft limit at six points. We therefore find that the 6-point effective Lagrangian can be written as
\eqs{
\frac{\mathcal{L}_{6}^{DBI}}{\sqrt{-g}}&=-\frac{1}{2}\nabla\phi\cdot\nabla\phi+\frac{d+1}{2}\phi^{2}-\frac{1}{8}(\nabla\phi\cdot\nabla\phi)^{2}+\frac{(d+1)(d+3)}{4!}\phi^{4}-\frac{1}{16}(\nabla\phi\cdot\nabla\phi)^{3}\\&\qquad\qquad+\frac{d+1}{16}(\nabla\phi\cdot\nabla\phi)^{2}\phi^{2}+\frac{2(d+1)(9-d^{2})}{6!}\phi^{6}.
\label{6pt1}
}
On the other hand, expanding the Lagrangian in \eqref{sgalLag} to sixth order (without applying equations of motion) gives
\eqs{
\frac{\mathcal{L}_{6}^{DBI}}{\sqrt{-g}}&=-\frac{1}{2}\nabla\phi\cdot\nabla\phi+\frac{d+1}{2}\phi^{2}-\frac{1}{8}(\nabla\phi\cdot\nabla\phi)^{2} - \frac{1}{4}(d+3)(\nabla\phi\cdot\nabla\phi)\phi^2\\
&\qquad\qquad + \frac{3(d+1)(d+3)}{4!}\phi^{4}-\frac{1}{16}(\nabla\phi\cdot\nabla\phi)^{3}-\frac{3(d+5)}{16}(\nabla\phi\cdot\nabla\phi)^{2}\phi^{2}\\
&\qquad\qquad-\frac{3(d+3)(d+5)}{48}(\nabla\phi\cdot\nabla\phi)\phi^{4}+\frac{15(d+1)(d+3)(d+5)}{6!}\phi^{6}.
\label{6pt2}
}
Matching the two Lagrangians using integration by parts and equations of motion is very tedious, so we instead verify that they give the same 6-point wavefunction coefficient in Appendix \ref{matching}.

\section{Conclusion}
\label{sec:conclusion}

In this paper, we have found evidence that the link between hidden symmetries and enhanced soft limits for scattering amplitudes in flat space extends to wavefunction coefficients in de Sitter space. In more detail, we have shown that enhanced soft limits fix the masses and couplings (including curvature corrections) of scalar effective field theories in agreement with the Lagrangians recently derived for the DBI and sGal theories from hidden symmetries in \cite{Bonifacio:2021mrf}. Moreover, we have shown that enhanced soft limits imply that the NLSM in dS must be massless and cannot receive curvature corrections, which would spoil its shift symmetry. We have carried out these calculations up to six points in the NLSM and DBI theory and four points in the sGal theory. At six points, the enhanced soft limits arise from cancellations between exchange and contact Witten diagrams, allowing us to fix all 6-point couplings in terms 4-point couplings. In principle, this procedure can be extended to any number of points allowing us to reconstruct the entire tree-level wavefunction coefficient, or equivalently the entire Lagrangian. 

There are a number of future directions. First of all, it would be interesting to extend our calculations to any number of points. This would involve writing down the most general effective action that reduces to the known one in the flat space limit, computing the tree-level wavefunction coefficients up to a given number of points using Witten diagrams, fixing the couplings from enhanced soft limits, and showing that the result agrees with the Lagrangians recently derived from hidden shift symmetries. If this were possible, it would be very significant because it would allow us to prove the relation between enhanced soft limits and hidden symmetries in dS. The difficulty with this approach is that the number of Witten diagrams quickly becomes very large at higher points. A more efficient method for fixing higher-point couplings from enhanced soft limits would therefore be very welcome. In flat space, enhanced soft limits allow one to define recursion relations for scattering amplitudes \cite{Cheung:2015ota,Bartsch:2022pyi}. It seems likely that similar progress can be made for wavefunction coefficients in dS by combining enhanced soft limits with knowledge of their singularity structure. This direction was recently explored in the context of flat space wavefunction coefficients, which do not exhibit enhanced soft limits but do obey soft theorems \cite{Bittermann:2022nfh}.

Another approach for fixing all couplings of the DBI theory from enhanced soft limits is suggested by the following observation. The DBI Lagrangian in dS can be written in the form
\begin{equation}
\frac{\mathcal{L}_{DBI}}{\sqrt{-g}}=\frac{\sqrt{1-X-Y}}{\left(1-Y\right)^{d/2+1}}=L(X,Y),
\end{equation}
where $X=\nabla\phi\cdot\nabla\phi$ and $Y=\phi^{2}$, which is a solution to the following simple differential equation:
\begin{equation}
\left(1-X-Y\right)\frac{\partial L}{\partial X}+\frac{L}{2}=0.
\end{equation}
In \cite{Cheung:2014dqa}, the flat space analogue of this differential equation (which corresponds to setting $Y=0$) was deduced from general arguments about enhanced soft limits of the S-matrix. Given the simplicity of the DBI Lagrangian in dS, it seems plausible that these arguments can be generalised to dS.

This leads us to the next question: how do we prove that higher shift symmetries in dS imply enhanced soft limits of the wavefunction coefficients without using Lagrangians? The analogous proof in flat space, which was sketched in section \ref{review}, relied heavily on the definition of the S-matrix, and does not immediately lift to wavefunction coefficients or CFT correlators. A useful strategy for addressing this question may be the one developed in \cite{Pimentel:2013gza,McFadden:2014nta} which studied soft limits of cosmological correlators from a boundary perspective. On the other hand, the shift symmetries underlying the NLSM, DBI, and sGal theories in dS are generated by diffeomorphisms which change the asymptotic behaviour of bulk fields so it is not immediately clear how to interpret them from the CFT perspective. We hope to gain a deeper understanding of this issue in the future.

Finally, it would be interesting to adapt our methods to other models. In flat space, \cite{Cachazo:2016njl,Cheung:2017ems} showed that soft limits of the NLSM, DBI, and sGal theories are actually controlled by larger theories which become visible when one expands beyond the order at which the soft limits vanish. It would therefore be interesting to extend our calculations to higher orders in the soft limit and investigate the emergence of extended theories in dS. Moreover the flat space Lagrangian for Born-Infeld theory (which is a vector effective field theory) can be uniquely fixed by the vanishing of multiple chiral soft limits \cite{Cheung:2018oki}, so it would natural to look for an analogue of this in dS. It would also be interesting to consider soft limits of more realistic inflationary models where Lorentz boosts are spontaneously broken \cite{Cheung:2007st,Green:2020ebl,Pajer:2020wxk}. In the flat space limit, the scattering amplitudes of such effective field theories do not generally exhibit enhanced soft limits, except in the case of the spontaneously broken DBI theory which exhibits an emergent Lorentz invariance with respect to the speed of sound \cite{Grall:2020ibl,Green:2022slj}. It would be fascinating if this phenomenon also occurs in de Sitter background.  

\begin{center}
\textbf{Acknowledgements}
\end{center}
We thank Humberto Gomez, Renann Lipinski Jusinskas, Paul McFadden, and Chia-Hsien Shen for useful discussions. CA and AL are supported by the Royal Society via a PhD studentship and a University Research Fellowship, respectively. JM is supported by a Durham-CSC Scholarship.

\appendix
\section{4-point sGal Soft Limit}\label{app:sGal}
This appendix includes some extra details of the calculations in section \ref{sgal4pt}. In particular, we will explain how to evaluate the $\hat{s}_{ab}^3$ terms in \eqref{4pteffectivesgal}. This is done using the definitions in \eqref{eqn:ConfGenerators} along with their known action on bulk-to-boundary propagators \cite{Armstrong:2022csc}:
\begin{equation}\label{eqn:ConfTimeGenerators}
\begin{array}{ccc}
D\mathcal{K}_{\nu} =  \eta\tfrac{\partial}{\partial\eta}\mathcal{K}_{\nu},  & &  P^{i}\mathcal{K}_{\nu} =  k^{i}\mathcal{K}_{\nu}, \\
 K_{i}\mathcal{K}_{\nu} =  \eta^{2}k_{i}\mathcal{K}_{\nu}, & & M_{ij}\mathcal{K}_{\nu} = 0.
\end{array}
\end{equation}
To evaluate the action of $\hat{s}_{ab}^3$ we also need 
\eqs{
K_\alpha(k_i\phi)&= \eta^2k_\alpha k_i\phi - 2\eta\delta_{i\alpha}\dot{\phi},\\
K_\alpha \dot{\phi} &= k_\alpha(\eta^2\dot{\phi} + 2\eta\phi),\\
K_\alpha(k_ik_j\phi)&= \eta^2k_\alpha k_ik_j\phi - 2(\delta_{\alpha i}k_j + \delta_{\alpha j}k_i)(\phi+\eta\dot{\phi}) + 2k_\alpha\delta_{ij}\phi,\\
K_\alpha \ddot{\phi} &= k_\alpha(\eta^2\ddot{\phi} + 4\eta\dot{\phi} + 2\phi),\\
D\dot{\phi} &= \eta\ddot{\phi} + \dot{\phi},\\
M_{12}\left(\vec{k}_1\cdot\vec{k}_2f(k_1,k_2)\right) & = 2(d-1)\vec{k}_1\cdot\vec{k}_2f(k_1,k_2),\\
M_{12}\left((\vec{k}_1\cdot\vec{k}_2)^2f(k_1,k_2)\right) & = 4\left(d(\vec{k}_1\cdot\vec{k}_2)^2 - k_1^2k_2^2\right)f(k_1,k_2),
}
where $f(k_1,k_2)$ is some function depending only on the magnitudes of the momenta. These operations are implemented in the code \verb|EnhancedSoftLimits.nb| in the form of replacement rules (this enables the integrands to be evaluated much faster than if it were evaluating them as derivatives). This code also has additional relations needed for 6-point calculations. 

The action of the cubic operator is then given by
\eqs{
\hat{s}_{12}^3\phi_1\phi_2 &= \eta^6\bigg[(\vec{k}_1\cdot\vec{k}_2)^3\phi_1\phi_2 + 3(\vec{k}_1\cdot\vec{k}_2)^2\dot{\phi}_1\dot{\phi}_2 + 3(\vec{k}_1\cdot\vec{k}_2)\ddot{\phi}_1\ddot{\phi}_2 + \dddot{\phi}_1\dddot{\phi}_2\\
&\qquad + \frac{3}{\eta}\Big(2(\vec{k}_1\cdot\vec{k}_2)^2(\dot{\phi}_1\phi_2 + \phi_1\dot{\phi}_2) + (\vec{k}_1\cdot\vec{k}_2)\left(-k_1^2\phi_1\dot{\phi}_2-k_2^2\dot{\phi}_1\phi_2 + 3(\ddot{\phi}_1\dot{\phi}_2 + \dot{\phi}_1\ddot{\phi}_2)\right)\\
&\qquad\qquad - k_1^2\dot{\phi}_1\ddot{\phi}_2 - k_2^2\ddot{\phi}_1\dot{\phi}_2 + \dddot{\phi}_1\ddot{\phi}_2+\ddot{\phi}_1\dddot{\phi}_2\Big)\\
&\qquad + \frac{1}{\eta^2}\bigg((10-3d)(\vec{k}_1\cdot\vec{k}_2)^2\phi_1\phi_2 + 2(\vec{k}_1\cdot\vec{k}_2)\Big(2(\ddot{\phi}_1\phi_2+\phi_1\ddot{\phi}_2) + (29-3d)\dot{\phi}_1\dot{\phi}_2\\
&\qquad\qquad-(k_1^2+k_2^2)\phi_1\phi_2\Big)+ 2k_1^2k_2^2\phi_1\phi_2\\
&\qquad\qquad-k_1^2(5\dot{\phi}_1\dot{\phi}_2+4\phi_1\ddot{\phi}_2)-k_2^2(5\dot{\phi}_1\dot{\phi}_2+4\ddot{\phi}_1\phi_2) + \dddot{\phi}_1\dot{\phi}_2+\dot{\phi}_1\dddot{\phi}_2\bigg)\\
&\qquad + \frac{1}{\eta^3}\left(4(3-d)\vec{k}_1\cdot\vec{k}_2(\dot{\phi}_1\phi_2+\phi_1\dot{\phi}_2)+(d-6)k_1^2\phi_1\dot{\phi}_2+(d-6)k_2^2\dot{\phi}_1\phi_2 + 3(\ddot{\phi}_1\dot{\phi}_2 + \dot{\phi}_1\ddot{\phi}_2)\right)\\
&\qquad + \frac{1}{\eta^4}\left((d-2)^2\vec{k}_1\cdot\vec{k}_2\phi_1\phi_2 + \dot{\phi}_1\dot{\phi}_2\right)\bigg].
}
We can then compute the soft limit:
\eqs{
\lim_{\vec{k}_{1}\rightarrow0}\hat{s}_{12}^{3}\phi_{1}\phi_{2} =& \eta^6\bigg(\dddot{\phi}_1\dddot{\phi}_2 + \frac{3}{\eta}\left(-k_2^2\ddot{\phi}_1\dot{\phi}_2+\dddot{\phi}_1\ddot{\phi}_2 + \ddot{\phi}_1\dddot{\phi}_2\right),\\
&\qquad + \frac{1}{\eta^2}\left(-k_2^2(5\dot{\phi}_1\dot{\phi}_2 + 4\ddot{\phi}_1\phi_1) + \dddot{\phi}_1\dot{\phi}_2 + 9\ddot{\phi}_1\ddot{\phi}_2 + \dot{\phi}_1\dddot{\phi}_2)\right),\\
&\qquad + \frac{1}{\eta^3}\left((d-6)k_2^2\dot{\phi}_1\dot{\phi}_2 + 3(\ddot{\phi}_1\dot{\phi}_2 + \dot{\phi}_1\ddot{\phi}_2)\right) + \frac{1}{\eta^4}\dot{\phi}_1\dot{\phi}_2\bigg) + \mathcal{O}(k_1).
}
This can also be expressed in terms of boundary generators as was done in previous work \cite{Armstrong:2022csc}. For example the leading soft limit of \eqref{4pteffectivesgal} is given by
\eqs{
\lim_{\vec{k}_{1}\rightarrow0}\Psi_{4}^{sGal} &= -\mathcal{N}\Big((\Delta-d)(\Delta-d-1)(\Delta-d-2)(D_2^3+D_3^3+D_4^3) \\
&\qquad\qquad -(\Delta-d)(\Delta-d-1)(B+2+2d)(D_2^2+D_3^2+D_4^2) \\
&\qquad\qquad+ \Delta(\Delta-d)(d(\Delta^2+\Delta-4)-B(2+\Delta)-\Delta^3+4\Delta-4) -C\Big)\\
&\qquad\qquad\qquad\int\frac{d\eta}{\eta^{\Delta+1}}\phi_2\phi_3\phi_4 + \mathcal{O}(k_1).
}

\section{6-point DBI Soft Limit} \label{6ptdbiappendix}

In this Appendix, we will provide more details about the calculation in section \ref{dbi6point}. In particular, we will present an algorithm for systematically applying equivalence relations to express the 6-point tree-level wavefunction coefficient in terms of linearly independent terms. This allows us to fix all the couplings from enhanced soft limits and is implemented in \verb|EnhancedSoftLimits.nb|. The equivalence relations are
\begin{itemize}
\item conformal Ward identities in terms of the $\hat{s}_{ab}$ operators,
\item boundary momentum conservation,
\item equations of motion for the bulk-to-boundary wavefunctions,
\item integration by parts identities/ addition of a total derivative to the integrand. 
\end{itemize}
Note that we neglect any boundary contributions that may come from integration by parts since they have delta function support when Fourier transformed to position space. Although the relations implied by conformal Ward identities can also be obtained from a combination of the other three equivalence relations, in practice we use all four in such a way as to remove the need for guesswork. In particular, we apply momentum conservation, equations of motion, and integration by parts relations in a particular order such that the latter can be constructed systematically.

After fixing $\Delta$ from the enhanced soft limit at four points, it is sufficient to work to leading order in the soft momentum in order to fix the 6-point couplings. The procedure for fixing these couplings is then given below:
\begin{enumerate}
\item Write the soft limit of an exchange diagram as a contact diagram by cancelling numerator and denominator in this limit (see \eqref{DBI6ptExchLim}).
\item Sum all diagrams over permutations to obtain the wavefunction coefficient. The wavefunction coefficient is now of the form $f(\hat{s}_{ab})\mathcal{C}_6$, where $f$ is a polynomial up to cubic order in the $\hat{s}_{ab}$.
\item Apply the conformal Ward identities to eliminate one leg and one $\hat{s}_{ab}$, mimicking the use of momentum conservation needed to demonstrate enhanced limits of amplitudes in flat space. We choose to eliminate leg $n$ and $\hat{s}_{n{-}2\,n{-}1}$ using $\hat{s}_{a n}=-\sum_{b=1}^{n-1}s_{a b}$ and $\left(\sum_{a=1}^{n-1}\mathcal{D}_a\right)^2=\hat{s}_{nn}$. At each stage we can also apply  $\hat{s}_{aa}\sim -m^2$. Note that this will remove any derivatives acting on the field $\phi_n$. It will not however remove all occurrences of $\vec{k}_{n{-}2}\cdot \vec{k}_{n{-}1}$ in the integrand since they can also appear from the successive action of $\hat{s}_{a \,n{-}2}\hat{s}_{a \,n{-}1}$, for example. This means that we can still apply boundary momentum conservation to eliminate quantities that are not independent. 
\item Use (\ref{eqn:SoftBtb}) to finish taking the soft limit and use the propagator equation of motion to remove factors of $k_a^2$.
\item Use boundary momentum conservation to remove $\vec{k}_{n{-}2}\cdot \vec{k}_{n{-}1}$. This will re-introduce the magnitudes $k_a^2$ (including $k_n$) so we again apply equations of motion such that the integrand contains only functions not linked by equations of motion.
\item The equations of motion will introduce derivatives of $\phi_n$ so use integration by parts to remove $\ddot{\phi}_n$ and then $\dot{\phi}_n$. This step can be done systematically by identifying terms of the form $\int d\eta\, g(\eta,\vec{k}_a,\partial_\eta^l\phi_{b\neq n})\partial_\eta^m\phi_n$ for some function $g$ and deriving the appropriate total derivative which contains it.
\item The wavefunction coefficient can now be seen to vanish for specific choices of the coefficients $A,B,C$ in \eqref{dbilagrangian}.
\end{enumerate}

Finally, we note that operators that are quadratic or cubic in leg 1 can be written as combinations of operators that are at most linear in leg 1, up to $\mathcal{O}(k_1)$. It is this property for example that allowed us to obtain equation \eqref{DBI6ptExchLim}. We also observe that
\eqs{
\hat{s}_{12}^3\mathcal{C}_{6}^{\Delta=d+1} &= \left((d^2+d+1)\hat{s}_{12}+d(d+1)\right)\mathcal{C}_{6}^{\Delta=d+1} + \mathcal{O}(k_1^2),\\
\hat{s}_{12}\hat{s}_{13}\hat{s}_{23}\mathcal{C}_{6}^{\Delta=d+1}&= \left(\hat{s}_{23}^2 + \hat{s}_{13}\hat{s}_{23} - (d+1)\hat{s}_{12} + d\hat{s}_{23}\right)\mathcal{C}_{6}^{\Delta=d+1} + \mathcal{O}(k_1^2).
}
Further demonstrations can be found in \verb|EnhancedSoftLimits.nb|. In principal, we could also use these properties to solve for the unknown coefficients without needing to consider the full integrand.

\section{Matching 6-point Wavefunctions} \label{matching}
We will now show that the wavefunction coefficient obtained from the Lagrangian in \eqref{6pt2} gives the same wavefunction coefficient as the one obtained from enhanced soft limits. Applying the free equation of motion to rewrite the $(\nabla\phi\cdot\nabla\phi)\phi^4$ as a $\phi^6$ interaction gives
\eqs{
\frac{\mathcal{L}_{6}^{DBI}}{\sqrt{-g}}&=-\frac{1}{2}\nabla\phi\cdot\nabla\phi+\frac{d+1}{2}\phi^{2}-\frac{1}{8}(\nabla\phi\cdot\nabla\phi)^{2}\\
&\qquad\qquad - \frac{1}{4}(d+3)(\nabla\phi\cdot\nabla\phi)\phi^2 + \frac{3(d+1)(d+3)}{4!}\phi^{4}-\frac{3}{48}(\nabla\phi\cdot\nabla\phi)^{3}\\
&\qquad\qquad-\frac{3(d+5)}{16}(\nabla\phi\cdot\nabla\phi)^{2}\phi^{2}+\frac{6(d+1)(d+3)(d+5)}{6!}\phi^{6}.
}
We then obtain the following contribution from 6-point contact Witten diagrams:
\eq{
\Psi^{DBI}_{6,\,\mathrm{cont}} = \delta^{3}\left(\vec{k}_{T}\right)\left[3\left(\hat{s}_{12}\hat{s}_{34}\hat{s}_{56}+\mathrm{perms}\right)-(5+d)\left(\hat{s}_{12}\hat{s}_{34}+\mathrm{perms}\right)+6(1+d)(3+d)(5+d)\right]\mathcal{C}_{6}^{\Delta=d+1},
\label{eqn:6ptRawCont}
}
where the terms are summed over all inequivalent permutations. Moreover, we find the following contribution from exchange diagrams: 
\eqs{
\Psi^{DBI}_{6,\,\mathrm{exch}}& = \frac{\delta^{3}\left(\vec{k}_{T}\right)}{(\mathcal{D}_{1}+\mathcal{D}_{2}+\mathcal{D}_{3})^{2}+m^{2}}\Big[\hat{s}_{12}\hat{s}_{3L}+\mathrm{Cyc.}[123]-3(1+d)(3+d)\\
&\qquad\qquad-(d+3)\left(\hat{s}_{12}+\hat{s}_{23}+\hat{s}_{31}+\mathcal{D}_{L}\cdot(\mathcal{D}_{1}+\mathcal{D}_{2}+\mathcal{D}_{3})\right)\Big]\times(123)\leftrightarrow(456)\mathcal{C}_{6}^{\Delta=d+1}+ \mathrm{perms}.
}

Next we use the conformal Ward identity at the vertex $-\mathcal{D}_L = \mathcal{D}_1+\mathcal{D}_2+\mathcal{D}_3$ to express the terms quadratic in boundary conformal generators terms as an inverse propagator plus a constant:
\eqs{
\Psi^{DBI}_{6,\,\mathrm{exch}}& = \frac{\delta^{3}\left(\vec{k}_{T}\right)}{(\mathcal{D}_{1}+\mathcal{D}_{2}+\mathcal{D}_{3})^{2}+m^{2}}\Big[\hat{s}_{12}\hat{s}_{3L}+\mathrm{Cyc.}[123]-3(1+d)(3+d)\\
&\qquad\qquad-(d+3)\left(\frac{1}{2}[(\mathcal{D}_{1}+\mathcal{D}_{2}+\mathcal{D}_{3})^{2}+m^{2}]+2(d+1)\right)\Big]\times(123)\leftrightarrow(456)\mathcal{C}_{6}^{\Delta=d+1}+\mathrm{perms},
}
where we have used $\mathcal{D}_a^2\sim-m^2$ to simplify the constant. This can be identified as the exchange diagram from \eqref{dbiexchange} plus a new contact contribution:
\begin{equation}
\Psi^{DBI}_{6,\,\mathrm{exch}} = \frac{\delta^{3}\left(\vec{k}_{T}\right)\Psi_{L}\Psi_{R}}{(\mathcal{D}_{1}+\mathcal{D}_{2}+\mathcal{D}_{3})+m^{2}}\mathcal{C}_{6}^{\Delta=d+1} + \tilde{\Psi}^{DBI}_{6,\,\mathrm{cont}},
\end{equation}
where
\begin{equation}
\tilde{\Psi}^{DBI}_{6,\,\mathrm{cont}}= \delta^{3}\left(\vec{k}_{T}\right)\left\{ \frac{1}{2}(d+3)(\Psi_{L}+\Psi_{R})+\frac{1}{4}(d+3)^{2}\left[(\mathcal{D}_{1}+\mathcal{D}_{2}+\mathcal{D}_{3})^{2}+m^{2}\right]\right\} \mathcal{C}_{6}^{\Delta=d+1}+\mathrm{perms}.\\
\end{equation}

We now work with the new contact contribution, summing over the 10 factorisation channels and comparing to the form in \eqref{eqn:6ptRawCont}. To do this, we want the quadratic term to be expressed as a sum of terms each with 4 distinct labels. We therefore use the conformal Ward identities to write $\mathcal{D}_L = \mathcal{D}_4+\mathcal{D}_5+\mathcal{D}_6$ to get
\eq{
\Psi_L = \hat{s}_{12}(\hat{s}_{34}+\hat{s}_{35}+\hat{s}_{36}) + \mathrm{Cyc}.[123] - (1+d)(3+d),
}
and analogously for $\Psi_R$. We can see that the quadratic term from $\Psi_L+\Psi_R$ will contain 18 terms so the sum over 10 channels will give a permutation-invariant sum of 180 terms. Since there are 45 unique $\hat{s}_{ab}\hat{s}_{cd}$, this gives us a symmetry factor of 4. A similar analysis of the linear terms from $(\mathcal{D}_1+\mathcal{D}_2+\mathcal{D}_3)^2$ gives a symmetry factor of 4 as well. We can therefore express the new contact contribution as
\eq{
\tilde{\Psi}^{DBI}_{6,\,\mathrm{cont}} = \delta^{3}\left(\vec{k}_{T}\right)\left[2(d+3)(\hat{s}_{12}\hat{s}_{34}+\mathrm{Perms})+(d+3)^{2}(\hat{s}_{12}+\mathrm{Perms})-5(d+1)(d+3)^{2}\right]\mathcal{C}_{6}^{\Delta=d+1}.
}
Noting that $(\hat{s}_{12} + \mathrm{perms}) = 3m^2=-3(d+1)$, this becomes
\eq{
\tilde{\Psi}^{DBI}_{6,\,\mathrm{cont}} = \delta^{3}\left(\vec{k}_{T}\right)\left[2(d+3)(\hat{s}_{12}\hat{s}_{34}+\mathrm{perms})-8(d+1)(d+3)^{2}\right]\mathcal{C}_{6}^{\Delta=d+1}.
}
We can then combine this with equation (\ref{eqn:6ptRawCont}) to give
\eq{
\Psi^{DBI}_{6,\,\mathrm{cont}} =\delta^{3}\left(\vec{k}_{T}\right)\left[(d+1)(\hat{s}_{12}\hat{s}_{34}+\mathrm{Perms})+2(d+1)(9-d^{2})\right]\mathcal{C}_{6}^{\Delta=d+1},
}
matching the result obtained from the enhanced soft limit. This wavefunction coefficient therefore also corresponds to the one obtained from \eqref{6pt1}.

%\bibliography{EffDoubleCopy}

\begin{thebibliography}{99}

\bibitem{Weinberg:1965nx}
S.~Weinberg,
``Infrared photons and gravitons,''
Phys. Rev. \textbf{140} (1965), B516-B524
doi:10.1103/PhysRev.140.B516

\bibitem{White:2011yy}
C.~D.~White,
``Factorization Properties of Soft Graviton Amplitudes,''
JHEP \textbf{05} (2011), 060
doi:10.1007/JHEP05(2011)060
[arXiv:1103.2981 [hep-th]].
 
\bibitem{Cachazo:2014fwa}
F.~Cachazo and A.~Strominger,
``Evidence for a New Soft Graviton Theorem,''
[arXiv:1404.4091 [hep-th]].

\bibitem{Strominger:2013jfa}
A.~Strominger,
``On BMS Invariance of Gravitational Scattering,''
JHEP \textbf{07} (2014), 152
doi:10.1007/JHEP07(2014)152
[arXiv:1312.2229 [hep-th]].

\bibitem{He:2014laa}
T.~He, V.~Lysov, P.~Mitra and A.~Strominger,
``BMS supertranslations and Weinberg\textquoteright{}s soft graviton theorem,''
JHEP \textbf{05} (2015), 151
doi:10.1007/JHEP05(2015)151
[arXiv:1401.7026 [hep-th]].

\bibitem{Arkani-Hamed:2008owk}
N.~Arkani-Hamed, F.~Cachazo and J.~Kaplan,
``What is the Simplest Quantum Field Theory?,''
JHEP \textbf{09} (2010), 016
doi:10.1007/JHEP09(2010)016
[arXiv:0808.1446 [hep-th]].

\bibitem{Gell-Mann:1960mvl}
M.~Gell-Mann and M.~Levy,
``The axial vector current in beta decay,''
Nuovo Cim. \textbf{16} (1960), 705
doi:10.1007/BF02859738

\bibitem{Weinberg:1968de}
S.~Weinberg,
``Nonlinear realizations of chiral symmetry,''
Phys. Rev. \textbf{166} (1968), 1568-1577
doi:10.1103/PhysRev.166.1568

\bibitem{Weinberg:1978kz}
S.~Weinberg,
``Phenomenological Lagrangians,''
Physica A \textbf{96} (1979) no.1-2, 327-340
doi:10.1016/0378-4371(79)90223-1

\bibitem{Adler:1964um}
S.~L.~Adler,
``Consistency conditions on the strong interactions implied by a partially conserved axial vector current,''
Phys. Rev. \textbf{137} (1965), B1022-B1033
doi:10.1103/PhysRev.137.B1022

\bibitem{Cheung:2016drk}
C.~Cheung, K.~Kampf, J.~Novotny, C.~H.~Shen and J.~Trnka,
``A Periodic Table of Effective Field Theories,''
JHEP \textbf{02} (2017), 020
doi:10.1007/JHEP02(2017)020
[arXiv:1611.03137 [hep-th]].

\bibitem{Cheung:2014dqa}
C.~Cheung, K.~Kampf, J.~Novotny and J.~Trnka,
``Effective Field Theories from Soft Limits of Scattering Amplitudes,''
Phys. Rev. Lett. \textbf{114} (2015) no.22, 221602
doi:10.1103/PhysRevLett.114.221602
[arXiv:1412.4095 [hep-th]].

\bibitem{Hinterbichler:2015pqa}
K.~Hinterbichler and A.~Joyce,
``Hidden symmetry of the Galileon,''
Phys. Rev. D \textbf{92} (2015) no.2, 023503
doi:10.1103/PhysRevD.92.023503
[arXiv:1501.07600 [hep-th]].

\bibitem{Maldacena:2002vr}
J.~M.~Maldacena,
``Non-Gaussian features of primordial fluctuations in single field inflationary models,''
JHEP \textbf{05} (2003), 013
doi:10.1088/1126-6708/2003/05/013
[arXiv:astro-ph/0210603 [astro-ph]].

\bibitem{Creminelli:2012ed}
P.~Creminelli, J.~Nore\~na and M.~Simonovi\'c,
``Conformal consistency relations for single-field inflation,''
JCAP \textbf{07} (2012), 052
doi:10.1088/1475-7516/2012/07/052
[arXiv:1203.4595 [hep-th]].

\bibitem{Hinterbichler:2013dpa}
K.~Hinterbichler, L.~Hui and J.~Khoury,
``An Infinite Set of Ward Identities for Adiabatic Modes in Cosmology,''
JCAP \textbf{01} (2014), 039
doi:10.1088/1475-7516/2014/01/039
[arXiv:1304.5527 [hep-th]].

\bibitem{Kundu:2014gxa}
N.~Kundu, A.~Shukla and S.~P.~Trivedi,
``Constraints from Conformal Symmetry on the Three Point Scalar Correlator in Inflation,''
JHEP \textbf{04} (2015), 061
doi:10.1007/JHEP04(2015)061
[arXiv:1410.2606 [hep-th]].

\bibitem{Creminelli:2003iq}
P.~Creminelli,
``On non-Gaussianities in single-field inflation,''
JCAP \textbf{10} (2003), 003
doi:10.1088/1475-7516/2003/10/003
[arXiv:astro-ph/0306122 [astro-ph]].

\bibitem{Assassi:2012zq}
V.~Assassi, D.~Baumann and D.~Green,
``On Soft Limits of Inflationary Correlation Functions,''
JCAP \textbf{11} (2012), 047
doi:10.1088/1475-7516/2012/11/047
[arXiv:1204.4207 [hep-th]].

\bibitem{Kundu:2015xta}
N.~Kundu, A.~Shukla and S.~P.~Trivedi,
``Ward Identities for Scale and Special Conformal Transformations in Inflation,''
JHEP \textbf{01} (2016), 046
doi:10.1007/JHEP01(2016)046
[arXiv:1507.06017 [hep-th]].

\bibitem{Shukla:2016bnu}
A.~Shukla, S.~P.~Trivedi and V.~Vishal,
``Symmetry constraints in inflation, $\alpha$-vacua, and the three point function,''
JHEP \textbf{12} (2016), 102
doi:10.1007/JHEP12(2016)102
[arXiv:1607.08636 [hep-th]].

\bibitem{Bonifacio:2021mrf}
J.~Bonifacio, K.~Hinterbichler, A.~Joyce and D.~Roest,
``Exceptional scalar theories in de Sitter space,''
JHEP \textbf{04} (2022), 128
doi:10.1007/JHEP04(2022)128
[arXiv:2112.12151 [hep-th]].

\bibitem{Ghosh:2014kba}
A.~Ghosh, N.~Kundu, S.~Raju and S.~P.~Trivedi,
``Conformal Invariance and the Four Point Scalar Correlator in Slow-Roll Inflation,''
JHEP \textbf{07} (2014), 011
doi:10.1007/JHEP07(2014)011
[arXiv:1401.1426 [hep-th]].

\bibitem{Maldacena:2011nz}
J.~M.~Maldacena and G.~L.~Pimentel,
``On graviton non-Gaussianities during inflation,''
JHEP \textbf{09} (2011), 045
doi:10.1007/JHEP09(2011)045
[arXiv:1104.2846 [hep-th]].

\bibitem{McFadden:2009fg}
P.~McFadden and K.~Skenderis,
``Holography for Cosmology,''
Phys. Rev. D \textbf{81} (2010), 021301
doi:10.1103/PhysRevD.81.021301
[arXiv:0907.5542 [hep-th]].

\bibitem{McFadden:2010vh}
P.~McFadden and K.~Skenderis,
``Holographic Non-Gaussianity,''
JCAP \textbf{05} (2011), 013
doi:10.1088/1475-7516/2011/05/013
[arXiv:1011.0452 [hep-th]].

\bibitem{Bittermann:2022nfh}
N.~Bittermann and A.~Joyce,
``Soft limits of the wavefunction in exceptional scalar theories,''
[arXiv:2203.05576 [hep-th]].

\bibitem{Raju:2012zr}
S.~Raju,
``New Recursion Relations and a Flat Space Limit for AdS/CFT Correlators,''
Phys. Rev. D \textbf{85} (2012), 126009
doi:10.1103/PhysRevD.85.126009
[arXiv:1201.6449 [hep-th]].

\bibitem{Bzowski:2013sza}
A.~Bzowski, P.~McFadden and K.~Skenderis,
``Implications of conformal invariance in momentum space,''
JHEP \textbf{03} (2014), 111
doi:10.1007/JHEP03(2014)111
[arXiv:1304.7760 [hep-th]].

\bibitem{Raju:2011mp}
S.~Raju,
``Recursion Relations for AdS/CFT Correlators,''
Phys. Rev. D \textbf{83} (2011), 126002
doi:10.1103/PhysRevD.83.126002
[arXiv:1102.4724 [hep-th]].

\bibitem{Arkani-Hamed:2015bza}
N.~Arkani-Hamed and J.~Maldacena,
``Cosmological Collider Physics,''
[arXiv:1503.08043 [hep-th]].
%421 citations counted in INSPIRE as of 28 Sep 2022

\bibitem{Arkani-Hamed:2017fdk}
N.~Arkani-Hamed, P.~Benincasa and A.~Postnikov,
``Cosmological Polytopes and the Wavefunction of the Universe,''
[arXiv:1709.02813 [hep-th]].

\bibitem{Arkani-Hamed:2018kmz}
N.~Arkani-Hamed, D.~Baumann, H.~Lee and G.~L.~Pimentel,
``The Cosmological Bootstrap: Inflationary Correlators from Symmetries and Singularities,''
JHEP \textbf{04} (2020), 105
doi:10.1007/JHEP04(2020)105
[arXiv:1811.00024 [hep-th]].

\bibitem{Sleight:2019hfp}
C.~Sleight and M.~Taronna,
``Bootstrapping Inflationary Correlators in Mellin Space,''
JHEP \textbf{02} (2020), 098
doi:10.1007/JHEP02(2020)098
[arXiv:1907.01143 [hep-th]].

\bibitem{Goodhew:2020hob}
H.~Goodhew, S.~Jazayeri and E.~Pajer,
``The Cosmological Optical Theorem,''
JCAP \textbf{04} (2021), 021
doi:10.1088/1475-7516/2021/04/021
[arXiv:2009.02898 [hep-th]].

\bibitem{Jazayeri:2022kjy}
S.~Jazayeri and S.~Renaux-Petel,
``Cosmological Bootstrap in Slow Motion,''
[arXiv:2205.10340 [hep-th]].

\bibitem{Pimentel:2022fsc}
G.~L.~Pimentel and D.~G.~Wang,
``Boostless Cosmological Collider Bootstrap,''
[arXiv:2205.00013 [hep-th]].

\bibitem{Pajer:2020wxk}
E.~Pajer,
``Building a Boostless Bootstrap for the Bispectrum,''
JCAP \textbf{01} (2021), 023
doi:10.1088/1475-7516/2021/01/023
[arXiv:2010.12818 [hep-th]].

\bibitem{Bzowski:2020kfw}
A.~Bzowski, P.~McFadden and K.~Skenderis,
``Conformal correlators as simplex integrals in momentum space,''
JHEP \textbf{01} (2021), 192
doi:10.1007/JHEP01(2021)192
[arXiv:2008.07543 [hep-th]].

\bibitem{Meltzer:2020qbr}
D.~Meltzer and A.~Sivaramakrishnan,
``CFT unitarity and the AdS Cutkosky rules,''
JHEP \textbf{11} (2020), 073
doi:10.1007/JHEP11(2020)073
[arXiv:2008.11730 [hep-th]].

\bibitem{Baumann:2020dch}
D.~Baumann, C.~Duaso Pueyo, A.~Joyce, H.~Lee and G.~L.~Pimentel,
``The Cosmological Bootstrap: Spinning Correlators from Symmetries and Factorization,''
SciPost Phys. \textbf{11} (2021), 071
doi:10.21468/SciPostPhys.11.3.071
[arXiv:2005.04234 [hep-th]].

\bibitem{Goodhew:2021oqg}
H.~Goodhew, S.~Jazayeri, M.~H.~Gordon Lee and E.~Pajer,
``Cutting cosmological correlators,''
JCAP \textbf{08} (2021), 003
doi:10.1088/1475-7516/2021/08/003
[arXiv:2104.06587 [hep-th]].

\bibitem{Meltzer:2021zin}
D.~Meltzer,
``The inflationary wavefunction from analyticity and factorization,''
JCAP \textbf{12} (2021) no.12, 018
doi:10.1088/1475-7516/2021/12/018
[arXiv:2107.10266 [hep-th]].

\bibitem{Sleight:2021plv}
C.~Sleight and M.~Taronna,
``From dS to AdS and back,''
JHEP \textbf{12} (2021), 074
doi:10.1007/JHEP12(2021)074
[arXiv:2109.02725 [hep-th]].

\bibitem{Bzowski:2022rlz}
A.~Bzowski, P.~McFadden and K.~Skenderis,
``A handbook of holographic 4-point functions,''
[arXiv:2207.02872 [hep-th]].

\bibitem{Armstrong:2022jsa}
C.~Armstrong, H.~Gomez, R.~Lipinski Jusinskas, A.~Lipstein and J.~Mei,
``New recursions for tree-level correlators in (Anti) de Sitter space,''
[arXiv:2209.02709 [hep-th]].

\bibitem{Roehrig:2020kck}
K.~Roehrig and D.~Skinner,
``Ambitwistor strings and the scattering equations on AdS$_{3}$\texttimes{}S$^{3}$,''
JHEP \textbf{02} (2022), 073
doi:10.1007/JHEP02(2022)073
[arXiv:2007.07234 [hep-th]]

\bibitem{Diwakar:2021juk}
P.~Diwakar, A.~Herderschee, R.~Roiban and F.~Teng,
``BCJ amplitude relations for Anti-de Sitter boundary correlators in embedding space,''
JHEP \textbf{10} (2021), 141
doi:10.1007/JHEP10(2021)141
[arXiv:2106.10822 [hep-th]].

\bibitem{Herderschee:2021jbi}
A.~Herderschee,
``A New Framework for Higher Loop Witten Diagrams,''
[arXiv:2112.08226 [hep-th]].

\bibitem{Armstrong:2022csc}
C.~Armstrong, H.~Gomez, R.~Lipinski Jusinskas, A.~Lipstein and J.~Mei,
``Effective field theories and cosmological scattering equations,''
JHEP \textbf{08} (2022), 054
doi:10.1007/JHEP08(2022)054
[arXiv:2204.08931 [hep-th]].

\bibitem{Eberhardt:2020ewh}
L.~Eberhardt, S.~Komatsu and S.~Mizera,
``Scattering equations in AdS: scalar correlators in arbitrary dimensions,''
JHEP \textbf{11} (2020), 158
doi:10.1007/JHEP11(2020)158
[arXiv:2007.06574 [hep-th]].

\bibitem{Gomez:2021qfd}
H.~Gomez, R.~L.~Jusinskas and A.~Lipstein,
``Cosmological Scattering Equations,''
Phys. Rev. Lett. \textbf{127} (2021) no.25, 251604
doi:10.1103/PhysRevLett.127.251604
[arXiv:2106.11903 [hep-th]].

\bibitem{Herderschee:2022ntr}
A.~Herderschee, R.~Roiban and F.~Teng,
``On the differential representation and color-kinematics duality of AdS boundary correlators,''
JHEP \textbf{05} (2022), 026
doi:10.1007/JHEP05(2022)026
[arXiv:2201.05067 [hep-th]].

\bibitem{Gomez:2021ujt}
H.~Gomez, R.~Lipinski Jusinskas and A.~Lipstein,
``Cosmological scattering equations at tree-level and one-loop,''
JHEP \textbf{07} (2022), 004
doi:10.1007/JHEP07(2022)004
[arXiv:2112.12695 [hep-th]].

\bibitem{Bern:2008qj}
Z.~Bern, J.~J.~M.~Carrasco and H.~Johansson,
``New Relations for Gauge-Theory Amplitudes,''
Phys. Rev. D \textbf{78} (2008), 085011
doi:10.1103/PhysRevD.78.085011
[arXiv:0805.3993 [hep-ph]].

\bibitem{Bern:2010ue}
Z.~Bern, J.~J.~M.~Carrasco and H.~Johansson,
``Perturbative Quantum Gravity as a Double Copy of Gauge Theory,''
Phys. Rev. Lett. \textbf{105} (2010), 061602
doi:10.1103/PhysRevLett.105.061602
[arXiv:1004.0476 [hep-th]].

\bibitem{Albayrak:2020fyp}
S.~Albayrak, S.~Kharel and D.~Meltzer,
``On duality of color and kinematics in (A)dS momentum space,''
JHEP \textbf{03} (2021), 249
doi:10.1007/JHEP03(2021)249
[arXiv:2012.10460 [hep-th]].

\bibitem{Armstrong:2020woi}
C.~Armstrong, A.~E.~Lipstein and J.~Mei,
``Color/kinematics duality in AdS$_{4}$,''
JHEP \textbf{02} (2021), 194
doi:10.1007/JHEP02(2021)194
[arXiv:2012.02059 [hep-th]].

\bibitem{Farrow:2018yni}
J.~A.~Farrow, A.~E.~Lipstein and P.~McFadden,
``Double copy structure of CFT correlators,''
JHEP \textbf{02} (2019), 130
doi:10.1007/JHEP02(2019)130
[arXiv:1812.11129 [hep-th]].

\bibitem{Lipstein:2019mpu}
A.~E.~Lipstein and P.~McFadden,
``Double copy structure and the flat space limit of conformal correlators in even dimensions,''
Phys. Rev. D \textbf{101} (2020) no.12, 125006
doi:10.1103/PhysRevD.101.125006
[arXiv:1912.10046 [hep-th]].

\bibitem{Alday:2021odx}
L.~F.~Alday, C.~Behan, P.~Ferrero and X.~Zhou,
``Gluon Scattering in AdS from CFT,''
JHEP \textbf{06} (2021), 020
doi:10.1007/JHEP06(2021)020
[arXiv:2103.15830 [hep-th]].

\bibitem{Jain:2021qcl}
S.~Jain, R.~R.~John, A.~Mehta, A.~A.~Nizami and A.~Suresh,
``Double copy structure of parity-violating CFT correlators,''
JHEP \textbf{07} (2021), 033
doi:10.1007/JHEP07(2021)033
[arXiv:2104.12803 [hep-th]].

\bibitem{Zhou:2021gnu}
X.~Zhou,
``Double Copy Relation in AdS Space,''
Phys. Rev. Lett. \textbf{127} (2021) no.14, 141601
doi:10.1103/PhysRevLett.127.141601
[arXiv:2106.07651 [hep-th]].

\bibitem{Drummond:2022dxd}
J.~M.~Drummond, R.~Glew and M.~Santagata,
``BCJ relations in ${AdS}_5 \times S^3$ and the double-trace spectrum of super gluons,''
[arXiv:2202.09837 [hep-th]].

\bibitem{Alday:2022lkk}
L.~F.~Alday, V.~Gon\c{c}alves and X.~Zhou,
``Supersymmetric Five-Point Gluon Amplitudes in AdS Space,''
Phys. Rev. Lett. \textbf{128} (2022) no.16, 161601
doi:10.1103/PhysRevLett.128.161601
[arXiv:2201.04422 [hep-th]].

\bibitem{Bissi:2022wuh}
A.~Bissi, G.~Fardelli, A.~Manenti and X.~Zhou,
``Spinning correlators in $\mathcal{N} = 2$ SCFTs: Superspace and AdS amplitudes,''
[arXiv:2209.01204 [hep-th]].

\bibitem{Sivaramakrishnan:2021srm}
A.~Sivaramakrishnan,
``Towards color-kinematics duality in generic spacetimes,''
JHEP \textbf{04} (2022), 036
doi:10.1007/JHEP04(2022)036
[arXiv:2110.15356 [hep-th]].

\bibitem{Cheung:2022pdk}
C.~Cheung, J.~Parra-Martinez and A.~Sivaramakrishnan,
``On-shell correlators and color-kinematics duality in curved symmetric spacetimes,''
JHEP \textbf{05} (2022), 027
doi:10.1007/JHEP05(2022)027
[arXiv:2201.05147 [hep-th]].

\bibitem{Cachazo:2014xea}
F.~Cachazo, S.~He and E.~Y.~Yuan,
``Scattering Equations and Matrices: From Einstein To Yang-Mills, DBI and NLSM,''
JHEP \textbf{07} (2015), 149
doi:10.1007/JHEP07(2015)149
[arXiv:1412.3479 [hep-th]].

\bibitem{Cachazo:2013hca}
F.~Cachazo, S.~He and E.~Y.~Yuan,
``Scattering of Massless Particles in Arbitrary Dimensions,''
Phys. Rev. Lett. \textbf{113} (2014) no.17, 171601
doi:10.1103/PhysRevLett.113.171601
[arXiv:1307.2199 [hep-th]].

\bibitem{Mason:2013sva}
L.~Mason and D.~Skinner,
``Ambitwistor strings and the scattering equations,''
JHEP \textbf{07} (2014), 048
doi:10.1007/JHEP07(2014)048
[arXiv:1311.2564 [hep-th]].

\bibitem{Cheung:2015ota}
C.~Cheung, K.~Kampf, J.~Novotny, C.~H.~Shen and J.~Trnka,
``On-Shell Recursion Relations for Effective Field Theories,''
Phys. Rev. Lett. \textbf{116} (2016) no.4, 041601
doi:10.1103/PhysRevLett.116.041601
[arXiv:1509.03309 [hep-th]].

\bibitem{Bartsch:2022pyi}
C.~Bartsch, K.~Kampf and J.~Trnka,
``Recursion Relations for One-Loop Goldstone Boson Amplitudes,''
[arXiv:2206.04694 [hep-th]].

\bibitem{Pimentel:2013gza}
G.~L.~Pimentel,
``Inflationary Consistency Conditions from a Wavefunctional Perspective,''
JHEP \textbf{02} (2014), 124
doi:10.1007/JHEP02(2014)124
[arXiv:1309.1793 [hep-th]].

\bibitem{McFadden:2014nta}
P.~McFadden,
``Soft limits in holographic cosmology,''
JHEP \textbf{02} (2015), 053
doi:10.1007/JHEP02(2015)053
[arXiv:1412.1874 [hep-th]].

\bibitem{Cachazo:2016njl}
F.~Cachazo, P.~Cha and S.~Mizera,
``Extensions of Theories from Soft Limits,''
JHEP \textbf{06} (2016), 170
doi:10.1007/JHEP06(2016)170
[arXiv:1604.03893 [hep-th]].

\bibitem{Cheung:2017ems}
C.~Cheung, C.~H.~Shen and C.~Wen,
``Unifying Relations for Scattering Amplitudes,''
JHEP \textbf{02} (2018), 095
doi:10.1007/JHEP02(2018)095
[arXiv:1705.03025 [hep-th]].

\bibitem{Cheung:2018oki}
C.~Cheung, K.~Kampf, J.~Novotny, C.~H.~Shen, J.~Trnka and C.~Wen,
``Vector Effective Field Theories from Soft Limits,''
Phys. Rev. Lett. \textbf{120} (2018) no.26, 261602
doi:10.1103/PhysRevLett.120.261602
[arXiv:1801.01496 [hep-th]].

\bibitem{Cheung:2007st}
C.~Cheung, P.~Creminelli, A.~L.~Fitzpatrick, J.~Kaplan and L.~Senatore,
``The Effective Field Theory of Inflation,''
JHEP \textbf{03} (2008), 014
doi:10.1088/1126-6708/2008/03/014
[arXiv:0709.0293 [hep-th]].

\bibitem{Green:2020ebl}
D.~Green and E.~Pajer,
``On the Symmetries of Cosmological Perturbations,''
JCAP \textbf{09} (2020), 032
doi:10.1088/1475-7516/2020/09/032
[arXiv:2004.09587 [hep-th]].

\bibitem{Grall:2020ibl}
T.~Grall, S.~Jazayeri and D.~Stefanyszyn,
``The cosmological phonon: symmetries and amplitudes on sub-horizon scales,''
JHEP \textbf{11} (2020), 097
doi:10.1007/JHEP11(2020)097
[arXiv:2005.12937 [hep-th]].

\bibitem{Green:2022slj}
D.~Green, Y.~Huang and C.~H.~Shen,
``Inflationary Adler Conditions,''
[arXiv:2208.14544 [hep-th]].


\end{thebibliography}
%\bibliographystyle{JHEP}

\end{document}